\documentclass[aps,prd,onecolumn,nofootinbib,superscriptaddress,12pt]{revtex4}
\pdfoutput=1
\usepackage{float}
\usepackage{graphicx}%Include figure files
\usepackage{amsmath}
\usepackage{enumitem}
\usepackage{empheq}
\usepackage{amsfonts}
\usepackage{amsmath}
\usepackage{amssymb}
\usepackage{amssymb,ulem}
\makeatletter
\renewcommand{\@makefnmark}{%
  \hbox{\@textsuperscript{\normalfont\textcolor{blue}{\@thefnmark}}}
}
\usepackage{dcolumn}
\usepackage{amsmath,amssymb,mathrsfs}
\usepackage{slashed}
\usepackage{subfigure}
\usepackage{multirow}
\usepackage{csquotes}
\usepackage{MnSymbol,wasysym}
\usepackage{braket,diagbox}
\usepackage{eurosym}
\usepackage[usenames,dvipsnames,svgnames]{xcolor}

\newcommand{\RNum}[1]{\uppercase\expandafter{\romannumeral #1\relax}}
\usepackage[colorlinks=true,linkcolor=blue,urlcolor=blue,filecolor=black,citecolor=red,
pdfstartview=FitV,pdftitle={},pdfsubject={},pdfkeywords={},pdfpagemode=None,bookmarksopen=true]{hyperref}
\newcommand{\const}{\mathrm{const}}

\usepackage{pifont}
\usepackage{booktabs}

\begin{document}
\baselineskip=0.5 cm

\title{Wiggling boundary and corner edge modes in JT gravity with defects}

\author{Kang Liu}
\email{liukanyang@sina.com}
\affiliation{\baselineskip=0.4cm Center for Gravitation and Cosmology, College of Physical Science and Technology, Yangzhou University, Yangzhou, 225009, China}

\author{Shoupan Liu}
\email{shoupan\_liu@163.com}
\affiliation{\baselineskip=0.4cm Center for Gravitation and Cosmology, College of Physical Science and Technology, Yangzhou University, Yangzhou, 225009, China}

\author{Xiao-Mei Kuang}
\email{xmeikuang@yzu.edu.cn}
\affiliation{\baselineskip=0.4cm Center for Gravitation and Cosmology, College of Physical Science and Technology, Yangzhou University, Yangzhou, 225009, China}

\begin{abstract}
\baselineskip=0.45 cm
We study the gravitational edge modes (GrEMs) and gauge edge modes (GaEMs) in Jackiw-Teitelboim (JT) gravity on a wiggling boundary. The wiggling effect manifests as a series of spacetime topological and bulk constraints for both conical and wormhole defect solutions. For the conical defect solution, we employ the generalized Fefferman-Graham (F-G) gauge to extend the boundary action, allowing for non-constant temperature and horizon position. We find that the infrared behavior of this boundary action is determined by the local dynamics of the temperature and horizon. For the wormhole defect solution, the boundary action can, in special cases, be described by a field with variable mass subject to a constant external force. We classify this corner system as a first-class constrained system influenced by field decomposition, confirming that the physical degrees of freedom are determined by constraints from the wiggling boundary information. We find that GrEMs and GaEMs can be linked at the corners by imposing additional constraints. Additionally, we show that the ``parallelogram'' composed of corner variables exhibits discreteness under a unitary representation. Finally, we explore that information from extrinsic vectors can be packaged into the GaEMs via a Maurer-Cartan form, revealing the boundary degrees of freedom as two copies of the $\mathfrak{sl}(2,\mathbb{R})$ algebra. By separating pure gauge transformations, we identify the gluing condition for gauge invariance and the corresponding integrable charges.
\end{abstract}

\maketitle
\tableofcontents
\section{Introduction}
In the quest for quantum gravity, the holographic principle has emerged as a guiding idea, suggesting that information in the spacetime bulk can be encoded on its boundary. The presence of a boundary fundamentally alters the nature of gauge symmetry \cite{REGGE1974286, Balachandran:1993tm, Balachandran:1994up, Carlip:1994gy, Carlip:2005zn, Balachandran:1995qa}. While gauge transformations are typically used to study redundancies in the bulk. The redundancies which do not vanish at a boundary can change the physical state. GaEMs are a set of new degrees of freedom or fields explicitly introduced at the corners, which are a priori independent of the pull-back of bulk fields \cite{Donnelly:2016auv, Speranza:2017gxd, Geiller:2017whh, Freidel:2018fsk, Geiller:2019bti,Donnelly:2020teo, Jiang:2020cqo,Freidel:2020xyx, Freidel:2020svx, Freidel:2020ayo}. The purpose of introducing GaEMs is to restore gauge invariance and to decouple the notion of corner symmetry from that of gauge symmetry.  They were  innovatively treated as boundary observables and degrees of freedom, which were  well-characterized for Abelian Chern-Simons theory \cite{Wen:2004ym, Balachandran:1991dw, Tong:2016kpv}.

A consistent description in local holography \cite{Dittrich:2017hnl,Dittrich:2017rvb} requires reconciling the corner structure of fields in the bulk with their behavior at the corners, where, for instance, commuting bulk variables can give rise to non-commuting properties \cite{Cattaneo:2016zsq,Ashtekar:1998ak}. GaEMs are the necessary ingredient for this reconciliation, carrying their own symplectic potential and connecting to the bulk via a ``gluing condition'' that preserves overall gauge invariance. These works reveal the corner symmetry algebra. In addition to the expected diffeomorphism charges, tetrad gravity possesses corner Lorentz charges, whose generators form a local $\mathfrak{sl}(2,\mathbb{C})$ algebra. Another significant point of progress is the identification of a non-commutative corner metric \cite{Freidel:2015gpa,Freidel:2018pvm} as a key dynamical variable of the corner phase space. The components of this corner metric are found to satisfy a local $\mathfrak{sl}(2,\mathbb{R})$ Poisson algebra. This non-commutativity is a fundamental feature of the gravitational phase space at the boundary.

A  powerful realization of the holographic principle is the AdS/CFT correspondence \cite{Maldacena:1997re, Gubser:1998bc, Witten:1998qj}, in which an illuminating and solvable instance is to connect the $(0+1)$-dimensional Sachdev-Ye-Kitaev (SYK) model to JT gravity \cite{Kitaev:15ur, sachdev1993gapless, Sachdev:2010um, Polchinski:2016xgd, you2017sachdev}. JT gravity is a rich theory of gravity coupled to a dilaton field in two-dimensional ${\rm AdS}$ spacetime \cite{Teitelboim:1983uy, Jackiw:1984je, Jensen:2016pah, Maldacena:2016upp}, and can be described by two types of actions. The first-order formulation action is a BF-like theory where the independent variables are a zero-form $B$-field and a spin-connection \cite{Chamseddine:1991fg}. In contrast, the second-order formulation is the original JT model, which utilizes a metric and a Lagrange multiplier field (a dilaton). The two actions can be converted into each other with the use of the spin-connection \cite{Frodden:2019ylc}. The SYK/JT correspondence provides a controlled environment to investigate black hole thermodynamics, quantum information, and the broken symmetries of holography \cite{Sachdev:2019bjn, Cvetic:2016eiv, Sachdev:2010um, Ozer:2025bpb}.

The symmetries of a horizon imply that there are degrees of freedom living at the boundary. These are associated with radial and boundary diffeomorphisms, which would normally be pure gauge deformations but are promoted to physical modes by the presence of boundary conditions \cite{Navarro-Salas:1998fgp, Carlip:2022fwh, Choi:2023syx}. This is a manifestation of the ``would-be gauge mode'' picture \cite{Carlip:2005tz}, which refers to transformations that represent pure symmetries in the bulk. At the boundary, however, these transformations no longer simply relate equivalent physical states. The parameters of the transformation themselves become dynamical degrees of freedom at the boundary, whose dynamical behavior can explain boundary physical phenomena, such as the entropy of the BTZ black hole \cite{Banados:1992wn}. {The GrEMs are additional physical degrees of freedom arising from gauge symmetries broken at the boundary, such as radial diffeomorphisms. They can be equivalently described as the wiggling boundary or a ``would-be gauge mode'' and its dynamics are governed by Schwarzian theory \cite{Joung:2023doq, Lee:2024etc}.}

Defects play a central role in JT gravity and can be understood from multiple perspectives. In the bulk, they can be described as point-like sources for the dilaton field \cite{ Witten:2020ert, Witten:2020wvy, Mefford:2020vde}. Adding point-like sources to the bulk action modifies the equations of motion. The partition function for this system, which describes a self-gravitating 0-brane, can be exactly solved as a Schwarzian theory, which is presented as an alternative to using a cosmic brane with tension to compute Rényi entropies \cite{Dong:2016fnf}. From a geometric viewpoint, these defects manifest as conical singularities (elliptic defects) or microscopic punctures, as well as macroscopic wormholes (hyperbolic defects) \cite{Mertens:2019tcm}. Alternatively, defects can be viewed as arising from corners at the boundary of spacetime, which are described by the Hayward term in the gravitational action \cite{Arias:2021ilh}. Such corners appear naturally in the computation of Hartle-Hawking wave functionals and reduced density matrices. This leads to the insertion of a defect operator into the partition function \cite{Jafferis:2019wkd}. Holographically, defects correspond to a deformation of the Schwarzian theory where the reparametrization mode is integrated over different coadjoint orbits of the Virasoro group \cite{Alekseev:1988ce,Stanford:2017thb}. For a study of corners in JT gravity with defects in the second-order formulation, see the references \cite{Arias:2021ilh, Botta-Cantcheff:2020ywu, Lin:2018xkj}.

Our research focuses on the codimension-one boundaries of solutions to second-order JT gravity, particularly those with conical and wormhole defects. The significant question in this scenario is how the wiggling boundary dictates the dynamics of GrEMs under given spacetime topological and bulk constraints, addressing the issues such as the boundary action from conical defects and the dynamics of the wormhole throat length. Furthermore, we investigate how GaEMs at the corners are influenced by this codimension-one boundary and its associated information. The significance of the corner stems from the fact that it precisely accounts for the difference between the first-order and second-order formulations of the theory.

This paper is organized as follows. In Section \ref{sec:Actions and internal symmteries}, we review JT gravity in both the first- and second-order formulations, and the explicit forms of the gauge transformations for the gauge connection and the dilaton field. In Section \ref{sec:Wiggling boundary with constraints}, specific radial and boundary diffeomorphisms are presented for spacetime with the  conical and wormhole defects, leading to the GrEMs of boundary dynamics. In Section \ref{edge modes from decompositions}, the relationship between the symplectic potentials in the first- and second-order formulation of JT gravity is derived. We classify corner canonical pairs, algebras, and observables, subject to constraints preserving the corner configuration and the wiggling boundary information. Using Maurer-Cartan form, we recast parts of external vectors as pure gauge, thereby achieving gauge invariance while ensuring non-trivial corner degrees of freedom. The last section contributes to our conclusions and discussion. Additionally, Appendix \ref{appe:appendix one} shows the notation and conventions adopted in our studies, and Appendix \ref{appendix: Examining the relationship between the first-order and second-order symplectic potential} presents some detailed calculations regarding the relation between the first- and second-order formulations symplectic potentials of JT gravity.

\section{First- and second-order formulations of the action}
\label{sec:Actions and internal symmteries}

The JT gravity circumvents the triviality of pure two-dimensional Einstein gravity by introducing a scalar field $X$, often interpreted as a dilaton, which acts as a Lagrange multiplier. The action of this theory in the second-order formulation is given by \cite{Teitelboim:1983ux,Jackiw:1982hg}
\begin{equation}
	\label{eq:JT_action}
	S=\int_{\mathcal{M}}d^{2}x\sqrt{\left| g \right|}X(R-\Lambda),
\end{equation}
where $g$ is the determinant of the two-dimensional metric $g_{\mu\nu}$, $R$ is the associated Ricci scalar, and $\Lambda$ is a cosmological constant.

An alternative, and in many ways more fundamental, description of JT gravity is available through a first-order, $\text{BF}$-like formulation. This approach recasts the theory in the language of gauge theory, where the fundamental variables are not metric and dilaton fields, but a gauge connection $A$  and a $B$-field, both valued in a Lie algebra. This action takes the form \cite{Chamseddine:1991fg}
\begin{equation}\label{eq:BF_action}
	S^F[B,A]=\int_{\mathcal{M}}\langle B,F(A)\rangle,
\end{equation}
where $F(A) = dA + A \wedge A$ is the curvature two-form of $A$, and $\langle \cdot, \cdot \rangle$ denotes a non-degenerate, invariant bilinear form on the chosen Lie algebra. The action \eqref{eq:BF_action} is invariant under the following transformations
\begin{subequations}\label{bulk field trans}
\begin{align}
&{\delta _\alpha }A =  - {\rm{d}}\alpha  - \left[ {A,\alpha } \right],\\
&{\delta _\alpha }{B} = \left[ \alpha ,{B} \right].
\end{align}
\end{subequations}

To establish the equivalence between the second-order JT action \eqref{eq:JT_action} and the first-order $\text{BF}$-like action described by its expanded form, one can choose the $\mathfrak{so}(2,1)$ algebra as a basis such that \eqref{eq:JT_action} is equivalent to \eqref{eq:BF_action}. This leads to the explicit form of the BF-like action
\begin{equation}
	\label{eq:BF_expanded}
	S^F[ B^a,B,{e^I},\omega ] = \int_{\cal M} {\left( {{\tilde B^a}\left( {{\rm d}{e_a} + {\epsilon_a}^b\omega  \wedge {e_b}} \right) + B{\rm d}\omega  + \frac{1}{2}B{\epsilon^{ab}}{e_a} \wedge {e_b}} \right)},
\end{equation}
 in which we have introduced the indices $a$ and indices $I$, such that $I = (a, X)$, where $X$ denotes a third type of index \cite{Grumiller:2017qao}. See Appendix \ref{appe:appendix one} for further details about the notation and convention used in this study. Note that the cosmological constant is hidden in the basis of the algebra \cite{Frodden:2019ylc}. Varying \eqref{eq:BF_expanded}  with respect to $\tilde B^a$ enforces the  equation of motion that the torsion two-form, defined as $T_a ={{\rm d}{e_a} + {\epsilon_a}^b\omega  \wedge {e_b}}$, must vanish. Here, $e$ and $\omega$ are zweibein and spin-connection,        respectively. Varying with respect to ${B}$ enforces the curvature constraint, ${{\rm d}\omega  + \frac{1}{2}{\epsilon^{ab}}{e_a} \wedge {e_b}}=0$. Under the assumption that the zweibein satisfies ${\nabla _\mu }{e^{\nu a}} = 0$ (see \cite{Oliveri:2019gvm} for a higher-dimensional example), the solution for the spin-connection is $\omega=\frac{1}{2}\epsilon_{ab}\omega^{ab}$, where $\omega^{ab}={\omega_\mu}^{ab}{\rm d} x^\mu=2{{e^\nu}^a\nabla_\mu{ e_\nu}^b}{\rm d} x^\mu$.
%The bridge from this first-order action back to the second-order JT action \eqref{eq:JT_action} is now straightforward. First, the torsion-free constraint, $T^a = 0$, allows one to solve for the spin connection $\omega$ algebraically in terms of the zweibein $e^a$. This solution renders $\omega$ a dependent variable, specifically the standard Levi-Civita spin connection, $\omega = \omega(e)$.
Substituting the $T_a=0$ condition back into the action \eqref{eq:BF_expanded}, the term multiplied by $B_a$ vanishes identically, as its constraint is now satisfied by construction. Then the action in the first-order formulation \eqref{eq:BF_expanded} can be reduced to
\begin{align}\label{effective action}
S^f = \int_{\mathcal{M}} {B} (d\omega(e) + \epsilon_{ab} e^a e^b),
\end{align}
in which the Lagrange multiplier ${B}$ now plays exactly the same role as the dilaton field $X$ in \eqref{eq:JT_action}. In the form of $S^f$, we can show that the first-order BF-like action \eqref{eq:BF_expanded} is indeed equivalent to the second-order JT action \eqref{eq:JT_action}.

Varying the action \eqref{eq:JT_action} with respect to $g_{\mu\nu}$ and $X$ respectively yields the equations of motion
\begin{subequations}
    \begin{align}
&g_{\mu \nu} X+\nabla_\mu \nabla_\nu X-g_{\mu \nu} \nabla^2X=0\label{dilaton field equation}, \\
&R+2=0\label{metric field equation}.
    \end{align}
\end{subequations}
Since we are interested in the transformation of the dilaton field at the boundary, which is manifested by the wiggling boundary, it is necessary to find the explicit form of the boundary action induced by diffeomorphism in the second-order formulation. These investigations will be carried out in the following sections. As for the first-order formulation, we will examine its corresponding gauge invariance and related issues in Section \ref{edge modes from decompositions}.

\section{Wiggling boundary with defects}
\label{sec:Wiggling boundary with constraints}
In this section, we shall study the wiggling boundary corresponding to the conical and wormhole defects of the JT gravity under radial and boundary diffeomorphisms. We will also investigate the boundary action to better understand the interplay between GrEMs and the wiggling boundary in solutions of JT gravity with defects. This analysis will focus on the geometric significance of the wiggling boundary within the generalized F-G gauge.

\subsection{Conical defect}
In order to study the wiggling boundary for conical defect, we should introduce the base spacetime coordinates $x^{b}=(\rho,t)$ which are built upon the radial and boundary diffeomorphisms of the target spacetime coordinates $x^t=(r,\tau)$. Further introducing functions ${\cal L}^{\pm,0}$, which represent the degrees of freedom in asymptotic $\rm{AdS}_2$ \cite{Grumiller:2017qao} and $\rm{AdS}_3$ \cite{Grumiller:2016pqb} theories, we can formulate  the zweibein $e$ and the spin-connection $\omega$ as
\begin{align}
	\begin{aligned}\label{tetrad and spin connection1}
		e_{r{0}} & =0, & e_{r{1}} & =1, & \omega_{r} & =0, \\
		e_{\tau{0}} & =e^{r}\mathcal{L}^{+}-e^{-r}\mathcal{L}^{-}, & e_{\tau{1}} & =\mathcal{L}^{0}, & \omega_{\tau} & =e^{r}\mathcal{L}^{+}+e^{-r}\mathcal{L}^{-},
	\end{aligned}
\end{align}
where $\tau$ is imaginary time. We can use the generalized F-G gauge in the target coordinate $(r,\tau)$ via the relation $g_{\mu\nu}=\eta_{ab}{e_\mu}^a{e_\nu}^b$, such that the metric becomes
\begin{align}\label{target ads}
{\rm{d}}{s^2} = {\rm{d}}{r^2} + 2{{\cal L}^0}{\rm{d}}r{\rm{d}}\tau  + \left({{({{\cal L}^0})}^2}- {{{({e^r}{{\cal L}^ + } - {e^{ - r}}{{\cal L}^ - })}^2} } \right){\rm{d}}{\tau ^2},
\end{align}
which is asymptotically $\rm{AdS}_2$. The dilaton field is solved from \eqref{dilaton field equation} as
\begin{align}\label{dilaton field}
   X^c=e^r \mathcal{X}^{+}(\tau)+e^{-r} \mathcal{X}^{-}(\tau),
\end{align}
where $\mathcal{X}^{+}$ and $\mathcal{X}^{-}$ are two functions that depend on the gauge choice, and their selection will affect the calculation of the boundary action as we will show in Subsection \ref{subsec:Extrinsic geometric quantities}. In particular, ${\cal L}^ +$  characterizes the boundary information, which is also our motivation for choosing this gauge. The location of the Killing horizon $r_h$ satisfies $e^{r_h}=\frac{1}{2 \mathcal{L}^{+}}\left( \pm {\mathcal{L}}^0 \pm \sqrt{4 \mathcal{L}^{+} \mathcal{L}^{-}+\left({\mathcal{L}}^0\right)^2}\right)$. Then the corresponding Hawking temperature is given by $T^{r_h}=\frac{1}{{\pi}}\sqrt{\frac{1}{4}(\mathcal{L}^{0})^{2}+\mathcal{L}^{+}\mathcal{L}^{-}}$.

Now, we move into the $(\rho,t)$  coordinate system in the target space, and consider the  general form of the radial and boundary diffeomorphisms
\begin{subequations}\label{radial transformation}
\begin{align}
	&	r \to w_1(t)\rho+\ln (w(t))+\frac{w_2(t)}{\rho}+{\cal O}\left( {\frac{1}{\rho^2}} \right),\label{radial transformation 1}\\
	&	\tau  \to \theta (t)+\frac{\theta_1(t)}{\rho} +{\cal{O}}\left( {\frac{1}{\rho^2}} \right),
\end{align}
\end{subequations}
where $w$ is called the radial displacement function and $\theta$ is known as the GrEM corresponding to the wiggling boundary \cite{Joung:2023doq}. The radial and boundary  diffeomorphisms in \eqref{radial transformation} must preserve the boundary gauge, which implies that the metric \eqref{target ads} in the  $(\rho,t)$  coordinates  should be
\begin{align}\label{wiggling fg}
	{\rm{d}}{s^2} = {\rm{d}}{\rho ^2} + 2{g_{\rho t}}{\rm{d}}\rho {\rm{d}}t + {g_{tt }}{\rm{d}}{t^2},
\end{align}
referred as the metric in base spacetime with the location of the Killing horizon $\rho_h$. To order ${\cal O}\left( {\frac{1}{\rho}} \right)$, we obtain two constraints,
\begin{subequations}\label{important condition 11}
\begin{align}
&w_1 =  1,\\
& w_2' =  - {{\cal L}^0}\theta_1'\label{fh relation},
\end{align}
\end{subequations}
which can ensure that $g_{\rho \rho}=1$, and
\begin{align}\label{cross constraint}
g_{\rho t}=\frac{w'}{w} + {{\cal L}^0}\theta '.
\end{align}
The prime here denotes a derivative with respect to $t$. Additional constraint on $g_{tt}$ depends on the form of conical defect, which will be studied soon.

%In what follows, we will simply denote $w_2$  as $w$.  Additionally, we will integrate further constraints in spacetime with defects.

\subsubsection{Conical defect on wiggling boundary}
\label{subsec:Conical defect on wiggling boundary}

Bulk defects in JT gravity are modeled as point-like sources for a dilaton field, $X$, which is accomplished by introducing a coupling term to the action of the form ${I_d } = 2\int_{\cal M} {{d^2}x \alpha (x)X(x)} $ with $\alpha^c(x)=\alpha^c\delta(x-x_0)$ with $x_0$ representing the location of the source \cite{Mertens:2019tcm, Witten:2020ert, Witten:2020wvy}. In this case, adding ${I_d }$ into \eqref{eq:JT_action} yields the equations of motion for the geometry with conical defect $R(x)+2=2\alpha^c\delta^2(x-x_0)$. The constant $\alpha^c$, given by
\begin{align}\label{conical defect anble}
  \alpha^c = 2{ \pi}(1- \frac{T^{r_h}}{T^{\rho_h}}),
\end{align}
measures the magnitude of the defect angle of the conical defect in the base spacetime. Here, $T^{\rho_h}$ is the black hole temperature defined in the base spacetime, which should reflect the conical defect near the black hole horizon. The temperature  $T^{r_h}$ sets the range of the coordinates  $\tau, t \in \left[ 0, \frac{1}{T^{r_h}} \right]$. 

It is known that diffeomorphisms themselves do not alter the topological properties of spacetime. The way to make a conical defect appear is generally to consider beforehand that the theory possesses an $I_d$ action. However, this is not the only mechanism; the behavior of the boundary degrees of freedom can also modify the spacetime topology \cite{Mertens:2019tcm,Mefford:2020vde,Blau:2024owj,Forste:2021roo}.
The degrees of freedom that influence the boundary behavior in our framework are $w$ and $\theta$. We term these internal degrees of freedom because their selection does not impact the results within the target spacetime, including the horizon's position and the black hole's temperature. However, we find that an internal symmetry exists in the $(w,\theta)$ space and  different realizations of this symmetry can determine the boundary dynamics.

Now we proceed to determine the conical defect angle $\alpha^c$ in $(\rho, t)$ coordinates. Since the behavior of $g_{tt}$ at leading order in $\rho$ is required to match that of $g_{\tau\tau}$ near the asymptotic boundary and the horizon, we expand $g_{tt}$ in two ways under conditions \eqref{important condition 11}. The expansion of $g_{tt}$ near the asymptotic boundary is
 \begin{align}\label{gtt constraints0}
 \begin{aligned}
& {g_{tt}} \approx \frac{{(w')^2}}{{{w^2}}} + \frac{{2{{\cal L}_0}w'\theta '}}{w} + \left( {{({\cal L}^0) }^2 + 2{{\cal L}^- }{{\cal L}^+ }} \right)(\theta ')^2 - {e^{ - 2\rho  - \frac{{2{w_2}}}{\rho }}}\left( {\frac{{2({{\cal L}^- })^2\theta_1'\theta '}}{{\rho {w^2}}} + \frac{{{({\cal L}^- \theta')}^2}}{{{w^2}}}} \right)\\
 &- {e^{2\rho  + \frac{{2w_2}}{\rho }}}\left( {\frac{{2{({\cal L}^+w) }^2\theta_1'\theta '}}{\rho } + ({{\cal L}^+ }{w}\theta {'})^2} \right) + \frac{1}{\rho }\left( { - \frac{{2{{\cal L}_0}\theta_1'w'}}{w} + 2\left( {{{\cal L}_0}^2 + 2{{\cal L}^- }{{\cal L}^+ }} \right)\theta_1'\theta '} \right.\\
&\left. { + 2{{\cal L}_0}\left( {\frac{{\theta_1'w'}}{w} - {{\cal L}_0}\theta_1'\theta '} \right)} \right)+{\cal O}(\frac{1}{\rho^2}).
\end{aligned}
 \end{align}
To eliminate the terms containing ${e^{\frac{1}{\rho }}}$ and ${e^{\frac{-1}{\rho }}}$ in $g_{tt}$, we directly set $w_2$ and $\theta_1$ to be zero, which do not alter the asymptotically $\rm{AdS}_2$ condition and the constraint \eqref{fh relation}. The second expansion is to the first order in ${\cal O}\left( {{\rho-\rho_h}} \right)$:
\begin{align}\label{gtt constraints}
\begin{aligned}
&{g_{tt}} \approx \frac{{({w}{')^2}}}{{{w}^2}} + \frac{{{{2\cal L}^0}{w}'\theta '}}{{{w}}} + \left({{({{\cal L}^0})}^2}- {\frac{{{e^{ - 2{\rho_h}}}{{\left( {{e^{2{\rho_h}}}{{\cal L}^ + }{w^2} - {{\cal L}^ - }} \right)}^2}}}{{{w^2}}} } \right){\theta '^2}\\
& - \frac{{2{e^{ - 2{\rho_h}}}\left( {{e^{2{\rho_h}}}{{\cal L}^ + }{w^2} - {{\cal L}^ - }} \right)\left( {{e^{2{\rho_h}}}{{\cal L}^ + }{w^2} + {{\cal L}^ - }} \right){({\theta '})^2}}}{{{w^2}}}\left( {\rho  - {\rho_h}} \right) + {\cal O}\left( {{{\left( {\rho  - {\rho_h}} \right)}^2}} \right).
\end{aligned}
     \end{align}
Then the horizon and temperature of the black hole in base spacetime are evaluated as 
\begin{subequations}\label{location of horizon}
    \begin{align}
        &{e^{{\rho _h}}} = \frac{1}{{2{{\cal L}^ + }{w^2}\theta '}}\left( { \pm (w' + w{{\cal L}^0}\theta ') \pm \sqrt {{{(w' + w{{\cal L}^0}\theta ')}^2} + 4{{\cal L}^ + }{{\cal L}^ - }{w^2}{{(\theta ')}^2}} } \right),\label{location of horizon 1}\\
        &T^{\rho_h} =   \frac{{\left( {{e^{ - {\rho_h}}}{{\cal L}^ - } + {e^{{\rho_h}}}{{\cal L}^ + }{w^2}} \right)\theta '}}{2{\pi}w}.\label{eq-T-rhoh}
    \end{align}
\end{subequations}
Note that unlike those in the target spacetime,  here $e^{\rho_h}$ and $T^{\rho_h}$  are not considered to be constants and they are also wiggling. Subsequently, the  conical defect angle can be calculated as
\begin{align}
    \alpha^c=2\pi \left( {1 - \frac{{w({e^{{r _h}}}{{\cal L}^ + } + {e^{ - {r _h}}}{{\cal L}^ - })}}{{\theta '({e^{{\rho _h}}}{{\cal L}^ + }{w^2} + {e^{ - {\rho _h}}}{{\cal L}^ - })}}} \right).
\end{align}
For simplicity, we can set $\mathcal{L}^- = 0$\footnote{This actually corresponds to a special case of the loosest set of boundary conditions considered in \cite{Grumiller:2017qao}. The significance of the loosest set of boundary conditions is that it allows the leading order coefficients of both the metric and the dilaton to fluctuate at the boundary, yielding the richest set of asymptotic symmetries, namely an $\mathfrak{sl}(2)$ current algebra.} to reduce 
\begin{align}\label{defect constraints}
 \alpha^c=2\pi(1-  \frac{e^{\Delta^r}}{w\theta^{\prime}})=2\pi \left( {1 - \frac{{{e^{{r_h}}}{{\cal L}^ + }}}{{2\pi {T^{{\rho _h}}}}}} \right),
\end{align}
where we have introduced the horizon displacement $\Delta^r=r_h-\rho_h=\ln \left| {\frac{{{{\cal L}^0}{w^2}\theta '}}{{w' + w{{\cal L}^0}\theta '}}} \right|$. If $w\theta'=1$ and $r_h=\rho_h$, we recover the similar case studied in \cite{Joung:2023doq}, in which the base spacetime has no conical defect.

We move on to the symmetry in the internal space $(w,\theta)$. The internal transformation changes the conical defect angle in the base spacetime without altering the position of the horizon. For simplicity, we still set ${\cal L}^-=0$. We define both types of transformations to generate a new GrEM, $\theta$, by transforming it as follows,
\begin{align}\label{first type inter trans}
    \theta  \to \frac{{\cos (\gamma )\theta  + \sin (\gamma )}}{{ - \sin (\gamma )\theta  + \cos (\gamma )}}.
\end{align}
A reasonable GrEM should revert to $t$ when the wiggling effect vanishes, implying that in this case the condition $\left( {\frac{{\cos (\gamma )\theta  + \sin (\gamma )}}{{ - \sin (\gamma )\theta  + \cos (\gamma )}}} \right)'=1$ must be satisfied. Furthermore, since $\gamma$ is not necessarily to be a constant, invariance of $\alpha^c$  \eqref{defect constraints} leads to the following transformation of the radial displacement function and $\Delta^r$ as\footnote{Here, the invariance of $\Delta_r$ can necessarily be achieved by introducing the transformations in \eqref{location of horizon 1} associated with ${\cal L}^0$ and ${\cal L}^+$.}
\begin{align}\label{angle preserving trans}
    w\to w{\left( {\cos (\gamma)   - \sin (\gamma) \theta } \right)^2}\frac{{\theta '}}{{\theta ' + \gamma '\left( {{\theta ^2} + 1} \right)}},\quad\quad \Delta^r\to \Delta^r.
\end{align}
The above transformations are based on the assumptions $ - \sin (\gamma )\theta  + \cos \gamma  \ne 0$ and $\theta ' + \gamma '\left( {{\theta ^2} + 1} \right) \ne 0$. For the  condition of vanishing GrEMs ($\theta'=1$ when $\theta=t$) to be satisfied after the transformation, a constraint must be imposed on $\gamma$
\begin{align}\label{constraint of gamma}
    {\mkern 1mu} {(t\sin \gamma  + \cos \gamma )^2} + \gamma '{\mkern 1mu} ({t^2} + 1) = 1,
\end{align}
which ensures that the transformed $\theta$ satisfies the condition for vanishing GrEMs at the same value of $t$. Note that for a constant $\gamma$, we must have $\gamma = n\pi$ for $n\in \mathbb{Z}$. This corresponds to either the identity transformation $\theta \to \theta$ or a reflection $\theta \to -\theta$ for $\theta$ (in the latter case, $w$ must also transform to $-w$). To obtain a continuous $\gamma$ that depends on $t$ and $\theta$, one must solve \eqref{constraint of gamma}.

It is worthwhile to note that the transformation of $\theta$ in \eqref{first type inter trans} is an elliptic Möbius transformation, which lies on the coadjoint orbit $\mathcal{O}_{\text{elliptic}} \cong \mathrm{Diff}(S^1)/U(1)$ \cite{Mertens:2019tcm}. In the classification of monodromy, the conical singularity corresponds to elliptic monodromy, which is geometrically interpreted as a massive, point-like particle defect. We will see how these internal transformations in the $(w, \theta)$ spaces, particularly the first type of transformations classifies the action and the boundary dynamics.

\subsubsection{Boundary action with conical defect}
\label{subsec:Extrinsic geometric quantities}
\begin{figure}
    \centering
    \includegraphics[width=0.9\linewidth]{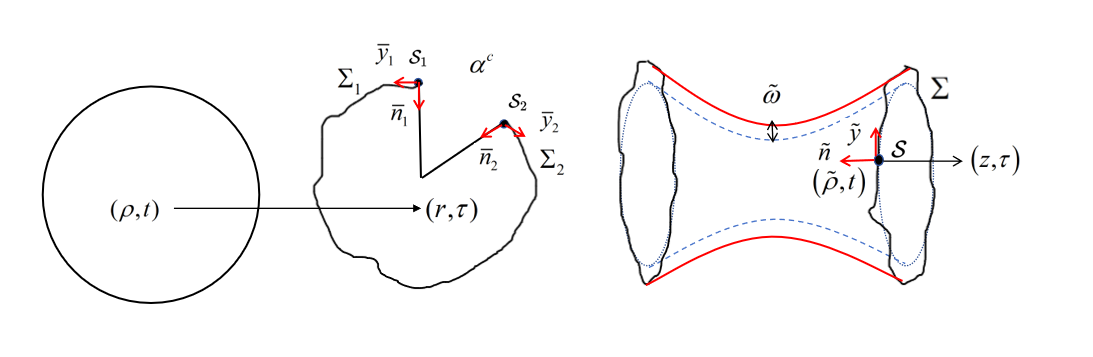}
    \caption {In the left panel, the normal and tangent vectors at the asymptotic boundary are marked with red lines with arrows, labeled $\bar y$ (which is one of $\bar y_1$ or $\bar y_2$) and $\bar n$ (which corresponds to the chosen $\bar y$), respectively. The conical defect angle is $\alpha^c$. Due to this angle, there are two sets of tangent and normal vectors, which intersect at the corners ${\cal S}_1$ and ${\cal S}_2$. The wiggling boundary corresponds to the curved boundary and is described by a coordinate transformation from base coordinates to target coordinates. In the right panel, we consider only a single corner, where a similar coordinate transformation exists. The normal and tangent vectors are labeled $\tilde n$ and $\tilde y$, respectively.}
    \label{Figure 1}
\end{figure}

After taking into account the constraints from the previous section, we can rewrite the metric of the base spacetime in $(\rho,t)$ coordinates
\begin{align}
    {\rm{d}}{s^2} = {\rm{d}}{\rho^2} +2 g_{\rho t }^0{\rm{d}}\rho{\rm{d}}t  + {g_{tt}}{\rm{d}}{t ^2},
\end{align}
where ${g_{tt}} = g_{tt}^ + {e^{2\rho }} + g_{tt}^ - {e^{ - 2\rho }} + g_{tt}^0$. We now compute the extrinsic geometric quantities in $\left( {\rho,t } \right)$ coordinates, which include the tangent vector $\bar y^\mu$ on $\Sigma$, its corresponding unit normal vector $\bar n^\mu$, and their associated extrinsic curvature $K^d$, as presented in the left panel of Fig. \ref{Figure 1}. To begin, $\bar y^\mu$ and $\bar n^\mu$ are given by
\begin{subequations}\label{extrinsic vectors 1}
\begin{align}
    &{\bar y^\mu } = \left( {\frac{{{\rm{d}}r}}{{{\rm{d}}t}},\frac{{{\rm{d}}\tau }}{{{\rm{d}}t}}} \right) ,\\
    &{\bar n^\mu } =\left(\bar n^\rho,\bar n^t \right),
\end{align}
\end{subequations}
where
\begin{subequations}
    \begin{align}
&{\bar n^\rho } = \frac{{g_{\rho t}^0{\bar y^\rho } + {\bar y^t}\left( {g_{tt}^ + {e^{2\rho }} + g_{tt}^ - {e^{ - 2\rho }} + g_{tt}^0} \right)}}{{\sqrt {\left( {{g_{tt}} - {{(g_{\rho t}^0)}^2}} \right)\left( {{{({\bar y^\rho })}^2} + 2g_{\rho t}^0{\bar y^\rho }{\bar y^t} + {{({\bar y^t})}^2}{g_{tt}}} \right)} }}\\
&{\bar n^t} = \frac{{{\bar y^\rho } + {\bar y^t}g_{\rho t}^0}}{{\sqrt {\left( {{g_{tt}} - {{(g_{\rho t}^0)}^2}} \right)\left( {{{({\bar y^\rho })}^2} + 2g_{\rho t}^0{\bar y^\rho }{\bar y^t} + {{({\bar y^t})}^2}{g_{tt}}} \right)} }}.
    \end{align}
\end{subequations}
As we focus on the asymptotic AdS$_2$ boundary, the normal vector is ${\bar n^\mu } =\left( { -1 + {\cal  O}({e^{ - 2\rho}}),{\cal  O}({e^{ - 2\rho}})} \right)$. The tangent vector is defined as the derivative with respect to $t$, which is the intrinsic parameter describing the boundary curve.
The extrinsic curvature in $(\rho,t)$ coordinate is
\begin{align}\label{extrinsic curvature 1}
 K^d &= - \frac{1}{2e^{2\rho}(g_{tt}^+ \bar y_2)^2} \Biggl[
  \bar y_2 (\bar y_1 + g_{r\tau}^0 \bar y_2) (g_{tt}^+)'  + g_{tt}^+ \biggl( \bar y_1^2 + \bar y_2 \Bigl[ (g_{r\tau}^0)^2 \bar y_2 - 2 \bigl( g_{tt}^0 \bar y_2 + \bar y_2 (g_{r\tau}^0)' + \bar y_1' \bigr) \Bigr] \nonumber \\
 &  + 2 \bar y_1 \bar y_2' \biggr) \Biggr] + \mathcal{O}(e^{-3\rho}). \nonumber
\end{align}
Ignoring the wiggling effect by setting $w=1$ and $\theta=t$, leads to $\bar y^\mu=(0,1)$. Then the corresponding extrinsic curvature $K_0^d$ is obtained by replacing {$\bar y_1=0$ and $\bar y_2=1$} in $K^d$. We choose $K_0^d$ as the counterterm such that the difference between $K^d$ and $K_0^d$ is finite:
\begin{align}
    K^d -{ K_0}^d = \frac{ g_{tt}^+ \left( 2\bar{y}_2 \bar{y}_1' - 2\bar{y}_1 \bar{y}_2' - \bar{y}_1^2 \right) - \bar{y}_1 \bar{y}_2 {g_{tt}^+}' }{ 2e^{2\rho} {g_{\tau \tau}^+}^2 \bar{y}_2^2 } + \mathcal{O}(e^{-3\rho}).
\end{align}
Note that the result in $K^d - {K_0^d}$ is unaffected by whether $g_{\tau\tau}^-$ is set to zero or not.

In the second-order formulation of JT gravity, an appropriate boundary term must be introduced to ensure that the variational problem is well-posed. We let $\Sigma$ be the wiggling boundary. Considering that \eqref{defect constraints} represents the constraint of the conical defect angle on the GrEM, $\theta$, we can use this relation to incorporate $\Delta^r$ and $T^{\rho_h}$ into the boundary action for the GrEM at $\rho=0$,
\begin{eqnarray}\label{boundary action main}
&&S_B = -2\int_{\Sigma} \sqrt{g_{\tau \tau}} X \mathrm{~d} t \left(K^d-K_0^d\right)  \nonumber \\
&&= -2\int_\Sigma {\rm{d}} t
       \left.\Big(- \frac{{Sch[\theta ,t]}}{{\theta '}} + \frac{1}{2e^{2\Delta^r} T^{\rho _h} (\theta ')^4}
       \Big[ - T^{\rho _h} (e^{\Delta^r}{\Delta^r}')^2 (\theta ')^2 (1 + 2\theta ') + 2e^{\Delta^r} T^{\rho _h} (\theta ')^3 (e^{\Delta^r})'' \right. \nonumber \\
&& + 2e^{\Delta^r} T^{\rho _h} (e^{\Delta^r})'(1 - \theta ')\theta '\theta '' + e^{2\Delta^r} T^{\rho _h} (\theta '')^2 (\theta ' - 1) + e^{\Delta^r} {T^{\rho _h}}'(\theta ')^2 \left( e^{\Delta^r}\theta '' - (e^{\Delta^r})'\theta '  \right) \Big] \Big)\nonumber \\
&&= - 2\int_\Sigma  {{\rm{d}}\theta \left.\Big( {Sch[t,\theta ] - \frac{{T_\theta ^{{\rho _h}}(\Delta _\theta ^r{t_\theta } + {t_{\theta \theta }})}}{{2{T^{{\rho _h}}}{t_\theta }}} - \frac{1}{2}{{(\Delta _\theta ^r)}^2}{t_\theta } + \Delta _{\theta \theta }^r - \Delta _\theta ^r{t_{\theta \theta }} - \frac{{t_{\theta \theta }^2}}{{2{t_\theta }}} + \frac{{t_{\theta \theta }^2}}{{2t_\theta ^2}}} \right.\Big)},
\end{eqnarray}
where the quantities with subscript $\theta$ indicate taking the first derivative with respect to $\theta$. Note that in the final equality, the integration variable has been changed from $t$ to $\theta$.  Here, the dilaton field \eqref{dilaton field} is chosen as ${{{\cal X}^ + }}=  - \frac{{{e^{{r_h}}}{{\left( {{e^{ - {r_h}}}\left( {2\pi  - \alpha } \right){T_{{\rho _h}}}} \right)}^{3/2}}}}{{{e^{{r_h}}}g_{tt}^ 0 + 2\left( {2\pi  - \alpha } \right){T_{{\rho _h}}}}}$ and with ${{{\cal X}^ - }}$ being arbitrary. The Schwarzian derivative, defined as $S{\rm{ch}}[\theta ,t] = \frac{{\theta '''}}{{\theta '}} - \frac{3}{2}\frac{{\theta '{'^2}}}{{\theta {'^2}}}$, also satisfies the useful identity $Sch[\theta ,t] = - \theta'^2 S{ch}[t,\theta ]$. Further using \eqref{defect constraints}, we can replace $\Delta^r$ in \eqref{boundary action main} with $\alpha^c$, which allows for the construction of an action that includes the conical defect angle. This approach is essentially equivalent to the original one because, as the following results will show, $\Delta^r$ and $T^{\rho_h}$ do not change the leading order of $S_B$.

To gain insight into this boundary action, we treat the wiggling effect perturbatively and perform the following expansion\footnote{To maintain generality, we assume that $\delta r$ and $\delta T^{\rho_h}$ only contain a subset of terms involving the derivative of $t$ with respect to $\theta$. This assumption is equivalent to the choice of ${\cal L}^+$ in \eqref{location of horizon}, which reflects the boundary information. This choice stems from the ambiguity in ${\cal L}^+$; we can expect that different constraints on ${\cal L}^+$ will produce different boundary dynamics. Here, we restrict our attention to the expansions that result in the subsequent second-order boundary action containing  certain polynomial in $\frac{2\pi n}{\beta}$.}:
\begin{subequations}\label{fucntions expansions}
\begin{alignat}{2}
    &t = \theta + \delta t(\theta),          &\qquad& \delta t =  \sum_n \varepsilon_n e^{\frac{2\pi in}{\beta}\theta}, \\
    &\Delta^r = \Delta ^{r(0)} + \delta r(\theta), &\qquad& \delta r = \alpha_0 \delta t + \alpha_1 \delta t_\theta + \alpha_2 \delta t_{\theta\theta}, \\
    &T^{\rho_h} = T^{\rho_h(0)} + \delta T^{\rho_h}(\theta), &\qquad& \delta T^{\rho_h} = \beta_0 \delta t + \beta_1 \delta t_\theta + \beta_2 \delta t_{\theta\theta},
\end{alignat}
\end{subequations}
in which the background fields $\Delta{^{r(0)}}$ and $T^{\rho _h(0)}$ are both independent of $\theta$, and we have already considered $\delta r$ and $\delta T^{\rho_h}$ to be only explicit functions of $t$. Our calculation shows that the zeroth-order term of $S_B$ is constant and the first-order is a total derivative, which means relations \eqref{location of horizon} and \eqref{defect constraints} are already on-shell. The second order of the action without expansion of $\varepsilon_n$ is calculated as
\begin{align}
     {\delta S_B^{(2)}} = \int d \theta \left( {{{(\delta {t_{\theta \theta }} + \delta {r_\theta })}^2} + \frac{{{{\delta {T^{\rho _h}_\theta}} }}}{{{T^{{\rho _h}(0)}}}}(\delta {t_{\theta \theta }} + \delta {r_\theta })} \right).
\end{align}
After considering the right panel of the expansion \eqref{fucntions expansions} and imposing periodic boundary conditions, the second-order term of the action is as follows
\begin{align}
 \delta S^{(2)} = \beta \sum_{n} \varepsilon_n \varepsilon_{-n} \mathcal{K}_n(P,Q) ,
\end{align}
in which
\begin{subequations}
    \begin{align}
& \mathcal{K}_n = P(k_n)P(k_{-n}) + \frac{1}{T^{\rho_h(0)}} P(k_n)Q(k_{-n}),\\
&P(k_n) = -i\alpha_2 k_n^3 - (1+\alpha_1)k_n^2 + i\alpha_0 k_n, \\
&Q(k_n) = -i\beta_2 k_n^3 - \beta_1 k_n^2 + i\beta_0 k_n,
\end{align}
\end{subequations}
and $k_n = \frac{2\pi n}{\beta}$. In this result, $P(k_n)$ and $Q(k_n)$ are functions representing the information of the horizon and temperature in base spacetime, which are constructed from the perturbation of $\delta t$. An instability, indicated by $\mathcal{K}_n < 0$, arises in the ultraviolet limit when the coefficient of the leading $k_n^6$ term  that depends on $\alpha_2$ and $\beta_2$ is negative. Further, in the infrared limit, we have
\begin{align}\label{IR}
    {{\cal K}_n} = \left( {\alpha _0^2 + \frac{{{\alpha _0}{\beta _0}}}{{{T^{{\rho _h}(0)}}}}} \right)k_n^2 + {\cal O}(k_n^4).
\end{align}
Note that the non-zero values of $\alpha_0$ and $\beta_0$ imply that the forms of $\Delta^r$ and $T^{\rho_h}$ must necessarily be $\Delta^r = \Delta^r(t,{t_\theta },{t_{\theta \theta }}, \cdots ,\int {\Gamma (t,{t_\theta },{t_{\theta \theta }}, \cdots )} d\theta )$ and ${T^{{\rho _h}}} = {T^{{\rho _h}}}(t,{t_\theta },{t_{\theta \theta }}, \cdots ,\int {\Gamma (t,{t_\theta },{t_{\theta \theta }}, \cdots )} d\theta )$. This means that the fields are local, because the conditions $\Delta^{r(0)} = \int {\delta (\theta  - \hat \theta ){\alpha _0}\delta t{\rm{d}}\hat \theta } $ and $T^{\rho_h(0)} = \int {\delta (\theta  - \hat \theta ){\beta _0}\delta t{\rm{d}}\hat \theta } $ hold. Therefore, the result \eqref{IR} implies that it is the local properties of both the horizon location changes and the temperature in the base spacetime that govern the behavior in the infrared limit.

{Additionally, according to the discussion below \eqref{constraint of gamma}, the transformations that preserve $\alpha$ also permit transformations with $w\to-w$ and $\theta\to-\theta$. Nonetheless, the boundary action \eqref{boundary action main} is generally not invariant under such transformations. In our study, achieving invariance of $\delta S_B^{(2)}$ under such transformations requires imposing constraints on the expansion coefficients. Since the terms that would violate this invariance are all those involving even-order derivatives with respect to $\theta$, we must substitute the expansions of $\delta r$ and $\delta T^{\rho_h}$ from \eqref{fucntions expansions} into $\delta S_B^{(2)}$ to isolate these parts and set the coefficients of these terms to zero. By doing so, we find that the condition for $\delta S_B^{(2)}$ to be invariant under transformations $w\to-w$ and $\theta\to-\theta$ can be ${\alpha _0} =  - \frac{{{\beta _0}}}{{{T^{{\rho _h}(0)}}}}$, ${\alpha _1} =  - 1$ and ${\alpha _2} =  - \frac{{{\beta _2}}}{{{T^{{\rho _h}(0)}}}}$. 

We now seek to understand the symmetries of the action $S_B$. We use the transformation law of the Schwarzian derivative under the composition of functions, given by the relation $Sch[{\cal F}(t),\theta ] = Sch[t,\theta ] + Sch[{\cal F}(t),t]t_\theta ^2$. This motivates us to rewrite the combination of the Schwarzian derivative and the final two terms from the last equality in \eqref{boundary action main} as follows:
\begin{align} \label{W term}
{\cal W}=Sch[t,\theta ] + \frac{{t_{\theta \theta }^2}}{{2{t_\theta }}} - \frac{{t_{\theta \theta }^2}}{{2t_\theta ^2}} = Sch[{\cal F}(t),\theta ] + c{{\cal F}'^2},
\end{align}
where $c$ is a constant. This relation can be shown to be equivalent to the Ermakov-Pinney equation \cite{leach2008ermakov}:
\begin{align}
    v(t)'' + \frac{{C(t)}}{2}v(t) = \frac{c}{2}v{(t)^{ - 3}},
\end{align}
in which $v = {\left( {{\cal F}'} \right)^2}$ and $C(t) = \frac{{t_{\theta \theta }^2}}{{2{t_\theta }}} - \frac{{t_{\theta \theta }^2}}{{2t_\theta ^2}}$\footnote{Note that the derivative with respect to $\theta$ can be converted back into a derivative with respect to $t$, which means that $\frac{{t_{\theta \theta }^2}}{{2{t_\theta }}} - \frac{{t_{\theta \theta }^2}}{{2t_\theta ^2}}$ can be expressed as a function of $t$.}. Under the assumption that $c$ is positive\footnote{The form of the solution for $\mathcal{F}$ depends on the value of $c$. When $c < 0$, $\mathcal{F}$ is a logarithmic function, while for $c = 0$, it becomes a rational function.}, the solution is given as follows
\begin{align}\label{F solution}
    {\cal F} = W\arctan (\xi ).
\end{align}
Here, we define $\xi = \frac{v_2}{v_1}$, where $v_1$ and $v_2$ are two linearly independent solutions to the equation $v''(t) + \frac{C(t)}{2}v(t) = 0$, which means that $\xi$, through $C(t)$, encapsulates the information of $\theta$ from the boundary action. The constant $W$ is their Wronskian, given by $W = v_1 v_2' - v_1' v_2= \pm \sqrt {\frac{c}{2}} $.
The motivation for constructing $\cal F$ is to enable a reparameterization of $t$ that restores the symmetry in the action, as was done in \cite{Arias:2021ilh,Joung:2023doq}. Based on \eqref{W term} and the form of $\cal F$ in \eqref{F solution}, the equivalent symmetry to keep $\cal W$ invariant is
\begin{align}
    \xi\to \frac{{\cos (\bar \gamma )\xi  + \sin (\bar \gamma )}}{{ - \sin (\bar \gamma )\xi  + \cos (\bar \gamma )}}.
\end{align}
Using this symmetry, the conical defect angle-preserving transformation \eqref{angle preserving trans} along with an identity transformation for $\theta$, and the form of $S_B$, we can write down the temperature transformation as
\begin{align}
    T^{\rho_h}\to \lambda T^{\rho_h},
\end{align}
where the parameter $\lambda$ is independent of $\theta$. This indicates that the boundary action possesses not only $\rm{SL}(2, \mathbb{R})$ symmetry but also rescaling symmetry of the temperature, provided that $\Delta^r$ remains unchanged. However, when $\Delta^r$ and $T^{\rho_h}$ in the base spacetime are constants, the terms in $S_B$ related to them will also vanish such that only the term $\cal W$ exists.

\subsection{Wormhole defect}
Another  defect solution to \eqref{metric field equation}, namely, a wormhole, takes the components of zweibein and spin-connection
\begin{align}
	\begin{aligned}\label{tetrad and spin connection2}
		{\tilde e_{z0}}  & =0, & {\tilde e_{z1}} & =  \frac{\varpi }{{{{{\mathop{\sin}\nolimits} }}\left( {\varpi z} \right)}}, & \tilde\omega_{z} & =0, \\
		\tilde e_{\tau{0}} & =  \frac{\varpi }{{{{{\mathop{\sin}\nolimits} }}\left( {\varpi z} \right)}}, &\tilde e_{z1} & =0, &\tilde \omega_{\tau} & = - \varpi \rm{ \cot}\left( {\varpi z} \right).
	\end{aligned}
\end{align}
The corresponding metric is
\begin{align}\label{wormhole solution}
    {\rm{d}}{s^2} = Z(z)({\rm{d}}{{\tau}^2} + {\rm{d}}{z^2}),
\end{align}
where $Z(\tau)=\frac{{{\varpi  ^2}}}{{{{\sin }^2}\varpi  z}}$. Different from the case in conical defect, here we choose an internal metric $\tilde \eta_{ab}$ given in \eqref{internal metric} due to convention. This solution has two boundaries: $z=0$ and $z=\frac{\pi}{\varpi}$, and the throat of the wormhole is located at $\frac{{\pi}}{2\varpi}$. The circumference of the throat is the length of the line element on the surface of constant $z$ at the position of the throat \cite{Mertens:2019tcm,Goel:2020yxl}. Assuming the range of $\tau$ in this case is now $\tau  \in \left[ {0,2{\pi}  } \right]$, it means that \eqref{wormhole solution} corresponds to a throat circumference of $l={\int {\varpi {\rm{d}}\tau } }=2{\pi}\varpi$. In this case, the solution for the dilaton field $X^w$ is
\begin{align}\label{eq:solution od dilation field2}
    X^w = \sqrt {Z(z)} \left( {\tilde C{e^{\varpi \tau }} + \tilde D{e^{ - \varpi \tau }}} \right)
\end{align}
where $\tilde C$ and $\tilde D$ are constants.  There is also a redundancy in the selection of $\tilde C$ and $\tilde D$, which is similar to the choice of ${\cal X}^+$ and ${\cal X}^-$ in \eqref{dilaton field}.

\subsubsection{Wormhole defect on wiggling boundary}
\label{subsec:wormhole }

To study the wiggling boundary, we consider the radial and boundary diffeomorphisms from the base coordinates $(\tilde \rho, t)$ to the target coordinates $(z, \tau)$. The line element in the base coordinates is written as
\begin{align}\label{base space in wormhole}
    {\rm{d}}{s^2} = {{\tilde g}_{\tilde \rho \tilde \rho }}{\rm{d}}{{\tilde \rho }^2} + 2{{\tilde g}_{\tilde \rho t}}{\rm{d}}\tilde \rho {\rm{d}}t + {{\tilde g}_{tt}}{\rm{d}}{t^2}.
\end{align}
Compared to \eqref{radial transformation}, we restrict our consideration to transformations near the boundary $z=0$ or equivalently $\tilde \rho =0$\footnote{In our expansion, we do not require $\tilde w_1$ to vanish at $\tilde \rho=0$, as we are considering an expansion around $z=0$.}:
\begin{subequations}\label{second type of trans}
    \begin{align}
& z \to {{\tilde w}_1}(t)+  {{\tilde w}}(t)\tilde\rho+ {{\tilde w}_2}(t)\tilde\rho^2 + {\cal O}({\tilde\rho^3}),\label{second type of trans1}\\
&\tau  \to {{\tilde \theta }}( t)+{\tilde \theta }_1(t)\tilde \rho+{\tilde \theta }_2(t)\tilde \rho^2 + {\cal O}({\tilde\rho^3}),
    \end{align}
\end{subequations}
where $\tilde \theta$ plays the role of the GrEM under wiggling effect. Similarly, to make sure that  $g_{\tilde \rho t}$ term is zero, and $g_{\tilde\rho\tilde\rho}$ and $g_{tt}$ are identical up to order $\mathcal{O}(\tilde\rho)$, we can constrain \eqref{second type of trans} into
\begin{subequations}\label{second main constarint}
\begin{align}
&\tilde w_1' = 0, ~~{{\tilde w}} =  \pm \tilde \theta ',~~{{\tilde w}_2} =\frac{1}{2}\varpi {\cot}\left( {\varpi \tilde w_1} \right){{\tilde w}}^2\\
&{{\tilde \theta }_1}' = \varpi \cot (\varpi {{\tilde w}_1})\tilde w\tilde \theta ',~~{{\tilde \theta }_2} =  - \frac{{{{\tilde w}}{{\tilde w}}'}}{{2\tilde \theta '}},\label{second main constarint final}
    \end{align}
\end{subequations}
under which, we have  ${{\tilde g}_{\tilde \rho \tilde \rho }}$, ${\tilde g}_{\tilde \rho t}$ and  ${{\tilde g}_{tt}}$ in $(\tilde \rho,t)$ coordinate as
\begin{align}
  \tilde Z=  {{\tilde g}_{\tilde \rho \tilde \rho }} = {{\tilde g}_{tt}}=\frac{({{\varpi }{{ {{{\tilde \omega }}} }}})^2}{{{{{\mathop{\sin}\nolimits} }^2}\left( {\varpi \tilde \omega_1 } \right)}}+{\cal O}\left( {\tilde \rho^2} \right),\quad {\tilde g}_{\tilde \rho t}=0+{\cal O}\left( {\tilde \rho^2} \right).
\end{align}
Additionally, the circumference of the throat in the base spacetime can be evaluated to be $l^N =2\pi \tilde w l$, which shows that $\tilde w$ is a factor measuring the change in the throat circumference, as illustrated in the right panel of Fig. \ref{Figure 1}.

\subsubsection{Boundary action with wormhole defect}
\label{subsec:Boundary action with wormhole defect}
To obtain the boundary action of the wormhole defect, we use a parallel approach in subsection \ref{subsec:Extrinsic geometric quantities}.  In this case, the tangent and normal vectors of the target spacetime in $\left( {\tilde\rho,t } \right)$ coordinates, corresponding to \eqref{wormhole solution}, are
\begin{subequations}
    \begin{align}
&{{\tilde y}^\mu } = \left( {{{\tilde y}_1},{{\tilde y}_2}} \right),\\
&{{\tilde n}^\mu } = \left( -{\frac{{{{\tilde y}_b}}}{H},  \frac{1}{H}} \right),
    \end{align}
\end{subequations}
where ${{\tilde y}_b} \equiv \frac{{{{\tilde y}_2}}}{{{{\tilde y}_1}}}$ and $H \equiv {\sqrt {\tilde Z\left( {1 + {{\tilde y}_b}^2} \right)} }$. Then, the extrinsic curvature $K^w$ in coordinate $\left( {\tilde \rho ,t} \right)$ is
\begin{align}
 K^w& = \frac{1}{2\tilde ZH(\tilde y_1^2 + \tilde y_2^2)} \Biggl[ 2\tilde y_1 \bigl[ \tilde y_2(\tilde y_b \tilde Z' - \partial_{\tilde \rho} \tilde Z) + \tilde Z(\tilde y_b \partial_{\tilde \rho} \tilde y_1 - \partial_{\tilde \rho} \tilde y_2) \bigr]  + 2\tilde y_2 \tilde Z (\tilde y_1' \tilde y_b - \tilde y_2') \nonumber \\
 & + (\tilde y_1^2 - \tilde y_2^2) (\tilde y_b \partial_{\tilde \rho} \tilde Z + \tilde Z') \Biggr].
\end{align}
When ignoring the wiggling effect, we have ${{\tilde y}^\mu } = \left( {\frac{{{\rm{d}}\tilde \rho }}{{{\rm{d}}t}},1} \right) = \left( {0,1} \right)$, such that the corresponding extrinsic curvature is
\begin{align}
  K^w_0= - \frac{{{\tilde y_b\partial _{\tilde \rho }}\tilde Z}}{{2{{\tilde Z}}}}.
\end{align}

 Considering the solution for the dilaton field  \eqref{eq:solution od dilation field2} with $\tilde C=\frac{1}{2}$ and $\tilde D=0$ and  choosing constraint $\tilde w=\theta'$ in \eqref{second main constarint final}, we directly perform the boundary action at $\tilde \rho=0$ for the wormhole solution as
\begin{align}\label{eq:boundary action in wormhole}
   {{\tilde S}_B} = \int_\Sigma  {{\rm{d}}t{e^{\varpi \tilde \theta }}\left( {S{\rm{c}}h[\tilde \theta (t),t] + \frac{{\tilde \theta ''}}{{\tilde \theta '}} + \frac{1}{2}{{\left( {\frac{{\tilde \theta ''}}{{\tilde \theta '}}} \right)}^2}} \right)},
\end{align}
which is a deformed Schwarzian theory. This expression contains an additional factor $e^{\varpi \tilde\theta}$ that cannot be eliminated by choosing a specific dilaton field, which is different from \eqref{boundary action main} for the conical defect. To more explicitly read off the boundary action of the GrEM, we introduce the variable $\phi (t) \equiv \ln \left| {\tilde \theta '(t)} \right|$, and rewrite \eqref{eq:boundary action in wormhole} as
    \begin{align}\label{eq:boundary action in wormhole1}
    {{\tilde S}_B} = \int_\Sigma  {{\rm{d}}t{\mkern 1mu} {e^{\varpi \mathop \smallint \nolimits^t {e^{\phi (t')}}{\rm{d}}t'}}\left( {\phi ''(t) + \phi '(t) + {{(\phi '(t))}^2}} \right)}.
\end{align}
In the absence of the wiggling effect, we have two interesting features: one is that the dilaton field becomes nearly time-independent, with its fluctuations approaching zero ($\phi \to 0$). The other is that the wormhole throat length undergoes slow, small-amplitude variations ($\varphi \to 0$). Consequently, we can expand \eqref{eq:boundary action in wormhole1} to first order in $\varpi$ as
    \begin{align}\label{approximate boundary action 1}
{{\tilde S}_B} \approx \int_\Sigma  {{\rm{d}}t\left( {1 + \varpi (t -{T^0})} \right)\left( {\phi '' + \phi ' + {{(\phi ')}^2}} \right)},
\end{align}
where ${T^0} =- \int_{}^t {{e^{\phi (\bar t)}}{\rm{d}}\bar t} $. By ignoring the terms on the co-dimension two surface, the above action admits the following equation of motion
\begin{align}
   \phi ''\left( {1 + \varpi (t - {T^0})} \right) + \varpi \phi ' + \frac{\varpi }{2} = 0,
\end{align}
from which we can be regarded $\phi$ as  a particle with variable mass. Here the coefficient  $2(1 + \varpi (t - T^0))$ corresponds to a mass that changes linearly with time, and $-\varpi$ corresponds to a constant external force. In principle, we can rewrite $S_B$ with $\phi$ as the integration variable, but this is merely a reparameterization and does not introduce any new physical properties, so we shall omit the formula. It is obvious that whether $\varpi$ is zero or not determines the existence of the wormhole, which can be understood as that the throat of the wormhole tends to disappear in target spacetime. This is reasonable because we have $Z = \frac{{{\varpi ^2}}}{{{\rm{Si}}{{\rm{n}}^2}\left( {\varpi z} \right)}}\mid_{\varpi \to 0} \to \frac{1}{{{z^2}}}$. Additionally, the boundary action \eqref{approximate boundary action 1} possesses a translational symmetry in $\phi$, meaning the action is invariant under the transformation $\phi\to \phi+\const$. However, due to the presence of the wormhole, i.e., $\varpi\ne0$, the symmetry in $t$ is lost. When the wormhole vanishes, both translation symmetries of the time and the GrEMs are simultaneously preserved.

Our results reveal that constraints on the spacetime topology and the bulk can influence the dynamical modes of GrEMs at the boundary. This influence is manifested in \eqref{radial transformation} and \eqref{base space in wormhole} and in the constraints on the quantities within them, which simultaneously reflects the breaking of diffeomorphism symmetry. Similar properties were also addressed in three-dimensional asymptotically AdS gravity \cite{Carlip:2005tz}, in which the dynamics of the non-trivial ``would-be gauge mode'' degrees of freedom at the boundary are described by Liouville theory. And the central charge, via the Cardy formula, precisely explains the microscopic origin of the BTZ black hole entropy. For a fixed boundary, a ``would-be gauge mode'' is a physical degree of freedom. If, however, one restores the broken symmetry by introducing a field, this field can subsequently be ``eaten'' by the coordinate transformations to become the GrEMs, at the cost of the boundary itself beginning to wiggle. Thus, the ``would-be gauge mode'' is revived as the dynamics of this wiggling boundary. In our case involving the asymptotic $\rm{AdS}_2$ and wormhole boundaries, we show that these ``eaten''  coordinates further reflect constraints from topology and bulk information, namely, the constant conical defect angle and the variation modes of the wormhole throat length.

\section{Corner algebra and gauge edge modes}
\label{edge modes from decompositions}
In this section, we investigate the GaEMs at corners. We aim to show how applying the configuration preserving conditions and the wiggling effect help us understand the corner. By appropriately defining canonical pairs, or corner variables, we can identify the algebras and observables. Eliminating the redundancies within these corner variables then allows for the identification of the Maurer-Cartan form and the classification of its symmetries. Ultimately, these considerations allow for the construction of a model of the GaEMs that preserves gauge invariance while possessing non-trivial charge.

\subsection{Relationship between first- and second-order formulation symplectic potentials}
\label{subsec:Relationship between first- and second-order formulation symplectic potentials}
We shall first demonstrate the relationship between the symplectic potentials, $\Theta^s$ and $\Theta^f$, which correspond to the actions \eqref{eq:JT_action} and \eqref{effective action}, respectively. The second- and first-order formulation of symplectic potentials are written as \cite{Goel:2020yxl,Arias:2021ilh}\footnote{According to \cite{Botta-Cantcheff:2020ywu,Jafferis:2019wkd,Arias:2021ilh}, corner terms present in the second-order formulation also contribute to the center of bulk. However, because they are typically determined by the boundary value of the dilaton field and are not among the corner variables we will consider, so they are excluded from our analysis here.}
\begin{subequations}
\begin{align}
    &\delta {S^s} \simeq {\Theta ^s}=\int_\Sigma  {  ( {2K\delta X  + {n_\mu }{\nabla ^\mu }X{h^{\alpha \nu }}\delta {h_{\alpha \nu }}})},\label{second order sym}\\
    &  \delta {S^f} \simeq {\Theta ^f} = \int_\Sigma  {\frac{1}{2}  B\varepsilon_{ab}\delta\omega^{ab} }\label{eq:first order symplectic},
\end{align}
\end{subequations}
where $\simeq$ means that the equations of motion are satisfied. Here $K$ is the extrinsic curvature with respect to its normal vector $n^\mu$. Definitions of differential forms, expressions for integrals in terms of differential forms, and the convention regarding the Levi-Civita symbol $\varepsilon_{ab}$ are presented in Appendix \ref{appe:appendix one}.

 According to the detailed calculations in Appendix \ref{appendix: Examining the relationship between the first-order and second-order symplectic potential}, we obtain an equivalence relation between $\Theta^s$ and $\Theta^f$ at the codimension-one boundary $\Sigma$:
\begin{align}\label{eq:second and first relation}
   {\Theta ^s} = {\Theta ^f} + \int_\Sigma  { {{\cal T}^{\alpha a}}\delta {e_{\alpha a}}}  + B{\left. {\left( {{n^\mu }{y_a }\delta {{e_\mu}^{ a}}} -y^a\delta n_a\right)} \right|_{\cal S}},
\end{align}
for $X={B}$ where $y^\mu$ is the vector tangent to $\Sigma$ but normal to the co-dimension-two surfaces $\cal S$ or corners (see Fig. \ref{Figure 1}), and ${{\cal T}^{\mu a}}$ is
\begin{align}\label{T tensor}
{{\cal T}^{\mu a}} = {n_\alpha }{\partial ^\alpha }{B}{e^{\mu a}} - {n^\mu }{\partial ^a}{B}.
\end{align}
This shows that in addition to the extra corner terms, a term containing ${{\cal T}^{\alpha a}}$ on $\Sigma$ must be subtracted from $\Theta^f$ so as to guarantee the equivalence of the two actions on $\Sigma$. The tensor ${{\cal T}^{\mu a}}$ is generally non-zero because the $B$-field is not a constant and its formula in general depends on the solutions to JT gravity. As we will show below, the term in \eqref{eq:second and first relation} containing ${{\cal T}^{\alpha a}}$ can be eliminated by redefining the $B$-field and the spin-connection $\omega$, provided that the corner terms acquire an additional term.

According to the actions \eqref{eq:JT_action} and \eqref{effective action}, the definitions of the $B$-field and the spin-connection $\omega$ admit the following ambiguity under a shift of total derivatives without changing the equations of motion\footnote{The ${\rm d} \omega_f$ term does not change the equations of motion, which is a consequence of the relation ${\rm d}^2=0$. The ambiguity in the choice of the spin-connection manifests itself through the appearance of the scalar $\omega_f$. This concept is also relevant to black hole thermodynamics \cite{Kastor:2009wy, Kastor:2008xb}.}
\begin{align}\label{extension of fields}
  B\to B + {B_r},\quad \omega\to\omega  + {\rm{d}}{\omega _r},
\end{align}
where $B+B_r$ satisfies equation of motion \eqref{dilaton field equation}. Subsequently, the symplectic potential of first-order formulation becomes
\begin{align}\label{eq:new symplectic potential relation}
   {\Theta ^f}\to{\Theta ^f} &= \int_\Sigma  {\left( {B + {B_r}} \right)} \delta \left( {\omega  + {\rm{d}}{\omega _r}} \right) \nonumber \\
   &= \int_\Sigma  {\left( {(B + {B_r})\delta \omega  - {\rm{d(}}B + {B_r}{\rm{)}}\delta {\omega _r}} \right)}  + {\left. {\left( {B + {B_r}} \right)\delta {\omega _r}} \right|_{\cal S}},
\end{align}
which means that the term containing ${{\cal T}^{\alpha a}}$ in \eqref{eq:second and first relation} will be eliminated  once  $B_r$ and $\omega_f$ satisfy the following matching  condition,
\begin{align}\label{eq:counter dondiction}
 \sqrt {\left| h \right|} {n_\mu }\epsilon^{\mu\nu}{\partial _\nu }\left( {B + {B_r}} \right)\delta {\omega _r} =  {{\cal T}^{\alpha a}}\delta {e_{\alpha a}},~~\text{with}~~ {{\cal T}^{\alpha a}}={n_\alpha }{\partial ^\alpha }{B}{e^{\mu a}} - {n^\mu }{\partial ^a}{B}.
\end{align}

We will then demonstrate that  a specific solution for $B_r$ and $\omega_f$ satisfying \eqref{eq:counter dondiction} indeed exists. Considering the metric  \eqref{target ads}, for the component $\delta e_{\tau 1}$ of $\delta {e_{\mu a}}$, one can choose $\omega_f=e_{\tau 1}$, while impose $B_r$ to simultaneously satisfy $\sqrt {\left| h \right|}{n_\mu }\epsilon^{\mu\nu}{\partial _\nu }\left( {B + {B_r}} \right) =  {{n_\mu }{\partial ^\mu }{B_r}{e^{\tau 1}} - {n^\tau }{e^{\mu 1}}{\partial _\mu }{B_r}}$. The indices $\bar 0$ and $\bar 1$ correspond to the abstract index. Since the relationship between $\partial_rB_{r}$ and $\partial_\tau B_{r}$ can be chosen arbitrarily, a solution for $B_r$ can always be found.  The parallel procedure can be applied to the other components of $\delta {e_{\mu a}}$. These components enable the introduction of further terms, similar to $\omega_f$ and $B_r$, following the form of \eqref{eq:new symplectic potential relation}. Then for the case where $w^{fr\bar0}, w^{fr\bar1},\cdots$ are equal to the other components of $ e^{\mu a}$, we can denote the corresponding solutions for $B_r^{r\bar0}, B_r^{\tau\bar1},\cdots$ sequentially as
\begin{align}
    \left( {\begin{array}{*{20}{c}}
{{w^{r\bar0}}}&{{w^{r\bar1}}}\\
{{w^{\tau\bar0}}}&{{w^{\tau\bar1}}}
\end{array}} \right) = \left( {\begin{array}{*{20}{c}}
{{e^{r\bar0}}}&{{e^{r\bar1}}}\\
{{e^{\tau\bar0}}}&{{e^{\tau\bar1}}}
\end{array}} \right) \Rightarrow \left( {\begin{array}{*{20}{c}}
{B_r^{r \bar0}}&{B_r^{r\bar1}}\\
{B_r^{\tau\bar0}}&{B_r^{\tau\bar1}}
\end{array}} \right) = B_r^{\mu a}.
\end{align}
 Therefore, the relation \eqref{eq:second and first relation} can be reduced to
\begin{align}\label{eq:new relation}
 {\Theta ^s}({\hat B})={\Theta ^f}(B) -{\Theta}^c,
\end{align}
where we introduce the field $\hat B=B+B_r^{r\bar0}+B_r^{r\bar1}+B_r^{\tau\bar0}+B_r^{\tau\bar1}$ satisfying the equation of motion \eqref{dilaton field equation}. Then the corresponding corner term is given by
\begin{align}\label{eq:corner term}
{\Theta}^c= -{\left. {\left( {{B}_r^{\mu a} +  {{n^\mu }{y_a }B\delta {{e_\mu}^{ a}}} -By^a\delta n_a} \right)\delta { e_{\mu a}}} \right|_{\cal S}} = {\left. {\left( {B{y^a}\delta {n_a} - {\bar B_r^{\mu a}}\delta {e_{\mu a}}} \right)} \right|_{\cal S}},
\end{align}
where we use the condition $\delta n^\mu \propto n^\mu$\footnote{The derivation here uses the property that the variation of the normal vector is proportional to the vector itself, i.e., $\delta n^\mu \propto n^\mu$. This is because, in this derivation, the position and shape of the boundary are assumed to be fixed. Similar treatment for JT gravity  can be found in \cite{Arias:2021ilh}. } and define ${\bar B_r^{\mu a}} = {B_r^{\mu a}} + B n^\mu y_a$. Note that $y^\mu$ is the normal vector to the corners  and is tangent to $n^\mu$.
In \cite{Arias:2021ilh}, the asymptotic corner terms in JT gravity were canceled by the chosen gauge and were consequently neglected. In our work, however, these corner terms are retained, as they are considered to be constrained by boundary information and to play an important role in subsequent gauge invariance.

As can be seen from \eqref{eq:new relation}, a non-zero value of ${B}_r^{\mu a}$ actually corresponds to the difference between the symplectic potentials of the first- and second-order formulations. If ${B}_r^{\mu a}=0$, then the integrals of $\Theta^s$ and $\Theta^f$ over $\Sigma$ are equal. Therefore, we name ${B}_r^{\mu a}$ as the \textit{residual tensor}. To better understand this tensor, we need to decompose generic Lorentz tensors and then generalize the decomposition of the $B$-field.

%Note that the residual dilaton field is not zero under the gauge transformation $\delta_\lambda$ introduced in \eqref{internal transformation}, i.e., $\delta_\lambda B_r\ne0$.

\subsection{Corner algebra}
\label{subsec:Corner algebra}
For a Lorentz tensor $A^{ab}=A\varepsilon^{ab}$, we can introduce its projection along $n_a$, denoted as $A_\bot^a = A^{ab}n_b$, such that it can be decomposed as
\begin{align}\label{original decom}
	{A^{ab}} = 2 A_ \bot ^{[a}{ n^{b]}}.
\end{align}
Henceforth, we will consistently use the identity matrix $\delta^{ab}=\tilde \eta^{ab}$ as the internal metric for abstract indices. In contrast to  the higher-dimensional \cite{Freidel:2020xyx,Freidel:2020svx,Freidel:2020ayo}, this decomposition lacks the dual part of $A^{ab}$ to introduce tangential components. This is because  the internal indices $a$, $b$ can only take two components due to the nature of the two-dimensional case. In order to understand how introducing ${ n}^a$ gives rise to tangential components at corners, we can redefine $B_r^{r\bar0}+B_r^{r\bar1}+B_r^{\tau\bar0}+B_r^{\tau\bar1}$ as $B_r$ and introduce the $B_r^{ab}$-field and $\tilde{\varepsilon}^a$:
\begin{align}
B_r^{ab}:=B_r\varepsilon^{ab},\quad	{{\tilde \varepsilon }^a}:= 2{\varepsilon ^{ab}}{{ n}_b}.
\end{align}\label{wg constrait ori}
The $B_r^{ab}$-field can be decomposed as
\begin{align}\label{eq:B field decom}
{B_r^{ab}}=2  \hat B^{[a} { n}^{b]}-\tilde{\varepsilon}^{[a} S^{b]},
\end{align}
where spin $S^a$ satisfying the constraint
\begin{align}\label{wg constrait}
	{S^{\bar 0}} = -\frac{{{B} +  {S^{\bar 1}}{n}^{\bar 1}}}{{ { n}^{\bar 0}}}\quad \text{or} \quad {B } =-\frac{1}{2} {\delta _{ab}}{S^{ a}}{ n^b}.
\end{align}
Here ${\hat B^{a}} = {\hat B}{\varepsilon ^{ab}}{{ n}_b}$ is treated as the projection of $B^{ab}$ onto ${ n}^{a}$, and $B_r$ was introduced in \eqref{extension of fields}.  From the decomposition in \eqref{eq:B field decom}, it appears that the Lorentz vector ${ n}^a$ merely provides the $B^{ab}$-field with a formal expansion on $\cal S$. The following discussion will show that $S^a$ provides a classification for the algebra at the corners, clarifying the significance of the radial part.

We now aim to identify the algebraic structure at the corner so that we can characterize the gauge symmetries present there. To proceed, we introduce $N^a$ by $N^a = B y^a$. For $(N^a,n^a)$ and $(\bar B^{\mu a},e^{\mu a})$ to be canonical pairs in \eqref{eq:corner term}, the following Poisson bracket $\{,\}_P$ must hold
\begin{subequations}\label{poinson brackets}
    \begin{align}
    &\{n_a, N^b\}_P= \delta_a^b ,\quad \{n_a, n^b\}_P=0,\\
    & \{N_a, N^b\}_P=0,\quad\{e_{\mu a}, \bar B^{\nu b}\}_P=- \delta_\mu^\nu \delta_a^b ,\\
    &\{e_{\mu a}, e^{\nu b}\}_P=0,\quad \{\bar B_{\mu a}, \bar B^{\nu b}\}_P=0.
    \end{align}
\end{subequations}
We shall restrict the corners to a fixed configuration that satisfies $N^an_a=0$ and $n^an_a=1$, which means that we are discussing a constrained configuration space. We define these two constraints as
\begin{subequations}
    \begin{align}
        \begin{aligned}
 &\phi_1= N^an_a\approx0,\\
&\phi_2=n_an^a-1\approx0,
\end{aligned}
    \end{align}
\end{subequations}
where $\approx$ denotes weak equality. These constraints imply that we are constraining the potential symmetries and algebraic structure at the corners to be within the set of configuration-preserving transformations. The constraints $\phi_1$ and $\phi_2$ can be compared those explicit solutions of constraints in \cite{Cianfrani:2008zv,Geiller:2011bh}. In addition to these two constraints, we also need to incorporate the results from Section \ref{sec:Wiggling boundary with constraints} concerning the codimension-one boundary. There, we were able to calculate the asymptotic behavior of two types of tangent vectors: ${y^\mu }{y_\mu } \approx g_{tt}^0{(\theta ')^2} + {\cal O}\left( {{\rho ^1}} \right)$ and ${{\tilde y}^\mu }{{\tilde y}_\mu } \approx {{\tilde g}_{tt}}{(\theta ')^2} + {\cal O}\left( {{{\tilde \rho }^2}} \right)$. These results indicate that the self-contraction of the boundary tangent vector encodes information about the GrEMs. Fundamentally, this shows that degrees of freedom from the codimension-one boundary also contribute to the results on the corner, implying that the corners are also wiggling. The corresponding constraint is
\begin{align}\label{phi3 cons}
    \phi_3= N^a N_a-V(\bar B^{\mu a},e^{\mu a},n^a)\approx0,
\end{align}
in which $V$ denotes the extension of constraints from the wiggling boundary information, such as those in \eqref{defect constraints} and \eqref{second main constarint}, onto the corners. We refer to $\phi_3$ as \textit{the constraint for the wiggling corner}, which represents the constraint imposed by the GrEMs on the GaEMs. Then according to \eqref{poinson brackets}, the constraint matrix for the constrained system is
\begin{align}
    M\approx
\begin{pmatrix}
\{\phi_1,\phi_1\}_P & \{\phi_1,\phi_2\}_P & \{\phi_1,\phi_3\} \\
\{\phi_2,\phi_1\}_P & \{\phi_2,\phi_2\}_P & \{\phi_2,\phi_3\} \\
\{\phi_3,\phi_1\}_P & \{\phi_3,\phi_2\}_P & \{\phi_3,\phi_3\}_P
\end{pmatrix}\approx
\begin{pmatrix}
0 & -2 & 2\tilde V \\
2 & 0 & 0 \\
-2\tilde V & 0 & 0
\end{pmatrix},
\end{align}
where $\tilde V = V + \frac{1}{2}{n_a}\frac{{\partial V}}{{{n_a}}}$. Since the first-class constraint is formed from the null vector of $M$, we designate the first-class constraint as $G$ and obtain
\begin{align}
    G = \tilde V{\phi _2} + {\phi _3}.
\end{align}
Here, $G$ generates gauge transformations of $e^{\mu a}, B^{\mu a}$, $n^a$ and $N^a$ that leave $\phi_1$, $\phi_2$ and $\phi_3$ invariant. Note that an observable is defined as a function of the canonical variables  whose Poisson bracket with all first-class constraints vanishes weakly.   The related Poisson brackets are calculated as
\begin{subequations}
\begin{align}
&{\left\{ {G,{n_a}} \right\}_P} \approx  - 2{N_a},\quad{\left\{ {G,{N^a}} \right\}_P} \approx 2\tilde V{n^a} - \frac{{\partial V}}{{\partial {n_a}}},\\
&{\left\{ {G,{e_{\mu a}}} \right\}_P} \approx  - \frac{{\partial V}}{{\partial {B^{\mu a}}}},\quad
{\left\{ {G,{\bar B^{\mu a}}} \right\}_P} \approx \frac{{\partial V}}{{\partial {e^{\mu a}}}}.
\end{align}
\end{subequations}
 Since the sub-matrix $\tilde M={\left( {{\tilde M^{ - 1}}} \right)^{ - 1}} = {\left( {\begin{array}{*{20}{c}}
0&{\frac{1}{2}}\\
{ -\frac{1}{2}}&0
\end{array}} \right)^{ - 1}}$ of $M$ is non-degenerate, it can be used to define the Dirac bracket:
\begin{align}
    {\left\{ {A_1,A_2} \right\}_D} = {\left\{ {A_1,A_2} \right\}_P} - \sum\limits_{i,j=1,2}\tilde M_{ij} {{{\left\{ {A_1,{\phi _i}} \right\}}_P}{{\left\{ {{\phi _j},A_2} \right\}}_P}}.
\end{align}
Subsequently, the bracket for the fields now changes from the Poisson bracket to the Dirac bracket\footnote{Unlike the method of handling constraints studied in \cite{Freidel:2020svx,Freidel:2015gpa,Donnelly:2016auv}, here we employ the method of Dirac bracket. This change in approach is motivated by the constraints arising from the GaEM via $\phi_3$, which corresponds to the boundary information for the wiggling boundary. Since first-class constraints are shown to exist in this setup, we consider this shift to the Dirac bracket formalism to be convenient. For similar work involving Dirac brackets, see \cite{Alexandrov:2000jw,Alexandrov:2002br}.}, ${\left\{ , \right\}_D}$:
\begin{subequations}
\label{eq:poisson_brackets_D_pairs}
\begin{alignat}{2}
    &\{e_{\mu a}, \bar B^{\nu b}\}_D = -\delta_{\mu}^{\nu} \delta_{a}^{b} ,
    &\quad
    &\{e_{\mu a}, e^{\nu b}\}_D = 0,
    \label{eq:poisson_eB_ee} \\
    &\{B_{\mu a}, \bar B^{\nu b}\}_D = 0,
    &\quad
    &\{n_a, N^b\}_D = \delta_a^b - n_a n^b,
    \label{eq:poisson_BB_nN} \\
    &\{n_a, n^b\}_D = 0,
    &\quad
    &\{N_a, N^b\}_D = n^b N_a-N^b n_a .
\end{alignat}
\end{subequations}
The $(e^{\mu a},\bar B^{\mu a})$ sector is now a well-defined subspace wherein an algebra can be constructed. We formally promote $e^{\mu a}$ and $\bar B^{\mu a}$ to ${\bf e}^{\mu a}$ and ${\bf B}^{\mu a}={\bf B}{{{\bf e}^{\mu}}_b}\varepsilon^{ba}$, respectively, representing the formal introduction of the corner variables. Furthermore, we can construct the transverse and longitudinal decompositions of ${{\bf e}^{\mu a}} = {{\bf e}^{T\mu }}{n^a} + {{\bf e}^{S\mu }}{\varepsilon _{ab}}{n^b}$ and ${{\bf B}^{\mu a}} = {{\bf B}^{T\mu }}{n^a} + {{\bf B}^{S\mu }}{\varepsilon ^{ab}}{n_b}$, where we introduce the following components for the tensor ${\bf W}^{\mu a}$:
\begin{align}
    \begin{aligned}
        & {{\bf W}^{T\mu }} = {{\bf W}^{\mu b}}{n_b},\\
        &{{\bf W}^{S\mu }} = {{\bf W}^{\mu b}}{\varepsilon _{bc}}{n^c}.
    \end{aligned}
\end{align}
The decomposition of ${\bf B}^{\mu a}$ can be directly compared with the result \eqref{eq:B field decom}; this only requires contracting both sides of \eqref{eq:B field decom} with one index of ${\bf e}^{\mu a}$. Further introducing $\boldsymbol{p} $,~$\boldsymbol{q}$
\begin{subequations}\label{eq:canonical_vars}
\begin{align}
    &\boldsymbol{q} = \arctan\left(\frac{n^{\bar 1}}{n^{\bar 0}}\right), \label{eq:q_def} \\
    &\boldsymbol{p} = n^{\bar 0} N^{\bar 1} - n^{\bar 1} N^{\bar 0}-2{\bf B}=\sqrt{V}-2\bf B, \label{eq:p_def}
\end{align}
\end{subequations}
helps to rewrite the corner symplectic potential  $\Theta^c$\eqref{eq:corner term} in the constraint space corresponding to $\phi_1, \phi_2,$ and $\phi_3$ as
\begin{align}\label{cononical sympl}
  \Theta^c\approx{\left. {\left( {\boldsymbol{p}\delta \boldsymbol{q} - {{\bf B}^{T\mu }}\delta {{\bf e}^T_{\mu a}}-{{\bf B}^{S \mu}}\delta {{\bf e}^S_{\mu }}} \right)} \right|_{\cal S}},
\end{align}
where we have combined the properties ${n_a}{n^a}-1  \approx 0$ and ${n^a}{n^b}{\varepsilon _{ab}} = 0$.

Using these brackets, we aim to further classify the symplectic potential of the form \eqref{cononical sympl} by constructing the generators of the ${\mathfrak{sl}}(2,\mathbb{R})$ algebra. The closed algebra satisfied by these generators should be independent of the presence of the constraints. Therefore, we must compute the Dirac brackets in order to construct the algebra.

We define the first set of generators as follows
\begin{align}
 j_{\bf e}^{(E)}=\frac{1}{2}g_{\mu\nu}{\bf e}^{\mu a}{\bf B}_a^\nu,\quad
 j_{\bf B}^{(E)}= \frac{1}{2}g_{\mu\nu}{\bf e}^{\mu a}{\bf e}_a^\nu,\quad
 j_{{\bf e}{\bf B}}^{(E)}=-\frac{1}{2}g_{\mu\nu}{\bf B}^{\mu a}{\bf B}_a^\nu.
\end{align}
Here, since ${\bf e}^{\mu a}$ has been promoted, we assume it no longer exists as a zweibein. In this case, the contraction of ${\bf e}^{\mu a}$ with itself does not yield a constant. Therefore, the algebra of $j_{\bf e}$ with the other generators does not vanish. The algebra they satisfy is
\begin{align}
\{j_{\bf e \bf B}^{(E)}, j_{\bf B}^{(E)}\}_D = -j_{{\bf B}}^{(E)},\quad
\{j_{{\bf eB}}^{(E)}, j_{\bf e}^{(E)}\}_D = j_{\bf e}^{(E)},\quad
\{j_{{\bf e}}^{(E)}, j_{\bf B}^{(E)}\}_D = 2j_{\bf e \bf B}^{(E)}.
\end{align}
 The second class of generators can be further and naturally divided into two types. We name these two types the $T$-type generators and the $S$-type generators, respectively. The $T$-type generators are
\begin{align}
    j_0^{(T)} = \frac{1}{2} g_{\mu\nu} {\bf e}^{T\mu} {\bf B}^{T\nu} ,\quad
    j_1^{(T)} = \frac{1}{2} g_{\mu\nu} {\bf e}^{T\mu}{\bf e}^{T\nu},\quad
    j_X^{(T)} = -\frac{1}{2} g_{\mu\nu} {\bf B}^{T\mu} {\bf B}^{T\nu},
    \end{align}
in which $k^T$ is an arbitrary constant and we have introduced two new fields ${\bf e}^{T\mu}={\bf e}^{\mu a}n_a$ and ${\bf B}^{T\mu}={\bf B}^{\mu a}n_a$. One can verify that they indeed satisfy the ${\mathfrak{sl}}(2,\mathbb{R})$ algebra
\begin{align}
\{j_0^{(T)},j_1^{(T)}\}_D=j_1^{(T)},\quad
\{j_0^{(T)},j_X^{(T)}\}_D=-j_X^{(T)},\quad
\{j_1^{(T)},j_X^{(T)}\}_D=2j_0^{(T)}.
\end{align}
 For the $S$-type, we have
\begin{align}
    j_0^{(S)}= \frac{1}{2} g_{\mu\nu} {\bf e}^{S\mu}{\bf B}^{B\nu} ,\quad
    j_1^{(S)} = \frac{1}{2} g_{\mu\nu} {\bf e}^{S\mu} {\bf e}^{S\nu},\quad
    j_X^{(S)} = -\frac{1}{2} g_{\mu\nu} {\bf B}^{S\mu}{\bf B}^{S\nu},
    \end{align}
which  satisfy
\begin{align}
\{j_0^{(S)},j_1^{(S)}\}_D=j_1^{(S)},\quad
\{j_0^{(S)},j_X^{(S)}\}_D=-j_X^{(S)},\quad
\{j_1^{(S)},j_X^{(S)}\}_D=2j_0^{(S)}.
\end{align}
Since the $(p, q)$ fields are decoupled from the $({\bf e}^{\mu a}, {\bf B}^{\mu a})$ fields, the Poisson brackets between all nine generators and both $\boldsymbol{p}$ and $\boldsymbol{q}$ vanish. The results for other types of Poisson brackets are as follows
\begin{subequations}\label{observable trans}
    \begin{align}
&{\left\{ {j_{\bf eB}^{(E)},{{\bf e}^{\mu a}}} \right\}_P} = {{\bf e}^{\mu a}},{\left\{ {j_{\bf e}^{(E)},{{\bf e}^{\mu a}}} \right\}_P} = 0,{\left\{ {j_{\bf B}^{(E)},{{\bf e}^{\mu a}}} \right\}_P} = 2{{\bf B}^{\mu a}}\\
&{\left\{ {j_{\bf eB}^{(E)},{{\bf B}^{\mu a}}} \right\}_P} =  - {{\bf B}^{\mu a}},{\left\{ {j_{\bf e}^{(E)},{{\bf B}^{\mu a}}} \right\}_P} =  - 2{{\bf e}^{\mu a}},{\left\{ {j_{\bf B}^{(E)},{{\bf B}^{\mu a}}} \right\}_P} = 0\\
&{\left\{ {j_0^{(T,S)},{{\bf e}^{\mu a}}} \right\}_P} = \frac{1}{2}\delta _{T,S}^a{{\bf e}^{\mu T,S}},{\left\{ {j_1^{(T,S)},{{\bf e}^{\mu a}}} \right\}_P} = 0,{\left\{ {j_X^{(T,S)},{{\bf e}^{\mu a}}} \right\}_P} =  - \delta _{T,S}^a{{\bf B}^{\mu T,S}}\\
&{\left\{ {j_0^{(T,S)},{{\bf B}^{\mu a}}} \right\}_P} =  - \frac{1}{2}\delta _T^a{{\bf B}^{\mu T}},{\left\{ {j_1^{(T,S)},{{\bf B}^{\mu a}}} \right\}_P} =  - \delta _{T,S}^a{{\bf e}^{\mu T,S}},{\left\{ {j_X^{(T,S)},{{\bf B}^{\mu a}}} \right\}_P} = 0,
    \end{align}
\end{subequations}
where $\delta_T^a=n^a$ and $\delta_S^a=n_b \varepsilon^{ab}$.

By employing the decomposition of ${\bf e}^{\mu a}$ and ${\bf B}^{\mu a}$, we can demonstrate the following relationship between the generators
\begin{align}
        j_I^{(E)} \approx j_I^{(E)} + j_I^{(S)}.
    \end{align}
Therefore, the Casimir operators corresponding to the generators $(j_{\bf e}^{(E)},j_{\bf B}^{(E)},j_{\bf BF}^{(E)})$, $(j_0^{(T)},j_1^{(T)},j_X^{(T)})$ and $(j_0^{(S)},j_1^{(S)},j_X^{(S)})$ are, respectively,
\begin{subequations}\label{generators}
    \begin{align}
&{{\cal C}_{\bf eB}} = \left( g_{\mu\nu}{\bf e}^{\mu a}{{\bf B}^{\nu }_a} \right)^2 - \left( g_{\mu\nu}{\bf B}^{\mu a}{{{\bf B}^{\nu }}_a} \right) \left( g_{\alpha\beta}{\bf e}^{\alpha b}{{\bf e}^{\beta } }_b\right),\\
&{{\cal C}^{T,S}} = \left( g_{\mu\nu} {\bf e}^{{T,S}\mu} {\bf B}^{{T,S}\nu} \right)^2 - \left( g_{\mu\nu} {\bf e}^{{T,S}\mu} {\bf e}^{{T,S}\nu} \right) \left( g_{\alpha\beta} {\bf B}^{{T,S}\alpha} {\bf B}^{{T,S}\beta} \right).
    \end{align}
\end{subequations}
 In fact, the value of the dilaton field on the horizon is proportional to the black hole entropy in dilaton gravity \cite{Gegenberg:1994pv}, {which gives the physical significance of the dilaton's value at the bulk center. This also indicates that for a unitary representation, which means the Casimir operators are characterized by the eigenvalues in the form $ {\cal C}_\lambda=\lambda \left( {\lambda  - 1} \right)$, where $\lambda$ is an integer. The Casimir operators in \eqref{generators}  that take discrete values can be interpreted as the square of the area of a``parallelogram'' formed by the ``sides'' ${\bf e}^{\mu a}$, ${\bf B}^{\mu a}$ and their $T, S$ components. This is because the form of the generators is analogous to the parallelogram's two sides, ${\bf l}_1$ and ${\bf l}_2$, which is precisely the square of the parallelogram's area: ${({{\bf l}_1} \times {{\bf l}_2})^2} = \left| {{{\bf l}_1}} \right|\left| {{{\bf l}_2}} \right| - {({{\bf l}_1} \cdot {{\bf l}_2})^2}$. Recall that ${\bf B}^{\mu a}$ relates to the residual tensor $B_r$ we constructed at the corners. Therefore, from the perspective of this ``parallelogram''-shaped Casimir operator on the asymptotic boundary, this residual tensor $B_r$ is effectively treated as ``another side'' of the ${\bf e}^{\mu a}$-field.}

{We now discuss how the function $V$, characterized by the wiggling boundary information, determines the observables.} Based on the decomposed symplectic potential \eqref{cononical sympl}, we have written the nine generators generically as $j$. It can be verified that the Poisson brackets $\{j, G\}_P$ all satisfy the following specific form, and from this result, the condition for the generators to be observables can be derived:
\begin{align}
    {\left\{ {j,G} \right\}_P} \approx  - {\left\{ {j,V} \right\}_P}=0 \Rightarrow V = V(j,{n^a}),
\end{align}
showing that the observables are determined by the specific form of $V$ introduced via the constraint $\phi_3$. For instance, if $V$ takes the form $V(j_{{\bf e}\bf B}, n^a)$, then ${\left\{ {{j_{{\bf e}\bf B}},G} \right\}_P} \approx  - {\left\{ {{j_{{\bf e}\bf B}},{j_{{\bf e}\bf B}}} \right\}_P}\frac{{\partial V}}{{\partial {j_{{\bf e}\bf B}}}} = 0$. Additionally, because the T-type and S-type generators commute (${\left\{ {{j^T},{j^S}} \right\}_P} = 0$), whereas the generators $j_{\bf{eB}}, j_{\bf e}$ and $j_{\bf B}$ do not commute with the S- and T-type generators, so $V$ can be written in a more specific form, 
\begin{align}\label{V form}
V=V(j^{(T)},j^{(S)},n^a),
\end{align}
where $j^{(T)}$ denotes a generator from the T-type algebra and $j^{(S)}$ denotes one from the S-type algebra. {The nine generators qualify as observables only when $V$ takes the form given in \eqref{V form}.} Moreover, given that one of the generators $j^{(E)}, j^{(T)}$ or $j^{(S)}$ is an observable, the condition for the Casimir operators to also be observables depends on the specific classification of $V$, such as the case $\left\{ {{\cal C},G} \right\}_P =  - \left\{ {{\cal C},V} \right\}_P =  - \left\{ {{\cal C},{j^{(E)}}}\right\}_P\frac{{\partial V}}{{\partial {j^{(E)}}}}$.

The purpose of decomposing $\Theta^c$ into canonical pairs can be understood from three points.  First, this approach avoids the ``danger around the corner" issue in \cite{Freidel:2020svx}, where the $B^{ab}$-field's corner components fail to commute (unlike the bulk field), by identifying them with the automatically commuting components of $e^{\mu a}$. The field $N^a$, serving as the conjugate momentum to $n^a$, encodes the combined the wiggling information of the GrEM and the dilaton field. This results in $\phi_3$, which describes the wiggling corners, being included in the constraints at the corners. Second, it permits a more detailed investigation of infinitesimal transformations. Although an alternative self-conjugate form for $\Theta^c$ like $s^\mu s_\mu$ by using the decomposition in \eqref{eq:B field decom}, it is less effective for constraining degrees of freedom, assigning physical meaning to the transformations. In addition, this division of conjugate pairs enables the recognition of GaEM transformations preserving physical and gauge invariants, along with the non-trivial degrees of freedom localized at corners. This will be shown in the subsequent discussion.

\subsection{Emergence of gauge edge mode}
\label{subsec:Emergence of gauge edge mode}
We will employ the covariant phase space formalism \cite{Wald:1993nt,Iyer:1994ys} to calculate the gauge charge and thereby establish the conditions for gauge invariance. The first step is to further rewrite the constrained expression for $\Theta^c$ from \eqref{cononical sympl}.

In order to isolate the gauge-invariant part of the symplectic potentials, we introduce a \textit{covariant variation}, ${\boldsymbol \delta}$. It is defined as the total variation, $\delta$, minus the pure gauge component, ${ \delta}_{\boldsymbol\lambda}$:
\begin{align}\label{total varia}
    {\boldsymbol \delta}:=\delta-{ \delta}_{\boldsymbol\lambda},
\end{align}
in which ${\boldsymbol\lambda}$ is a gauge matrix that generates the Lorentz transformation of ${\bf e}^{\mu a}$, given by ${\delta _{\boldsymbol\lambda} }{{\bf e} }^{\mu a} =\boldsymbol\lambda^{ab}{{\bf e}^{\mu}}_b$ \cite{Grumiller:2021cwg,Grumiller:2017qao}. It should be noted that this transformation is defined only for  ${\bf e}^{\mu a}$, whereas the transformations for $\boldsymbol{p}$ and $\boldsymbol{q}$ are generated by the generators obtained in the previous section.  We now consider the symplectic potential of the first-order formulation for GaEMs, obtained by rewriting \eqref{cononical sympl} exclusively in terms of the introduced corner variables $\left({{\bf B}^{T\mu a}}, {{{\bf e}^{\mu a}},{\boldsymbol \vartheta}^{ab}({\boldsymbol p}, {\boldsymbol q})}\right)$. Among these GaEMs, the new field ${\boldsymbol \vartheta}^{ab}$, should classify the gauge transformations of the corner variables, and it must also ensure the closure of the algebra satisfied by the gauge charges (see the further statements below). This requires that ${\boldsymbol\lambda}={\boldsymbol \vartheta}^{-1}\delta {\boldsymbol \vartheta}$ in \eqref{total varia}, which is called Maurer-Cartan form\footnote{This differential form embodies the profound connection between the local geometric structure of the Lie group as a manifold and its tangent space. A core characteristic of this form is its invariance under the left-multiplication action of groups, which is constitutes one of its most fundamental properties \cite{frankel2004geometry}.}.

 We select the matrix ${\boldsymbol\vartheta}^{ab}$ whose elements depend only on $\boldsymbol{q}$, and specify its form as follows
 \begin{align}
    &{\boldsymbol\vartheta}^{ab}  =
\begin{pmatrix}
&\cos(\boldsymbol{q}) & -\sin(\boldsymbol{q}) \\
&\sin(\boldsymbol{q}) & \cos(\boldsymbol{q})
\end{pmatrix},
\end{align}
which means that ${\boldsymbol\vartheta}^{ab}$ is an element of the $\mathrm{SO}(2)$ group. We can expect that different choices for $\boldsymbol{\vartheta}$  lead to different ${\Theta}^{\cal S}$\footnote{Note that the choice for $\vartheta^{ab} $ depends on which one-parameter subgroup of $\rm{SL}(2,\mathbb{R})$ is chosen. For example, we can choose $\rm{SO}(1,1)$: $\boldsymbol{\vartheta}^{ab}  = \left( {\begin{array}{*{20}{c}}
{\cosh (q)}&{\sinh (q)}\\
{\sinh (q)}&{\sinh (q)}
\end{array}} \right)$, or we can choose null rotations: $\boldsymbol{\vartheta}^{ab}  = \left( {\begin{array}{*{20}{c}}
1&q\\
0&1
\end{array}} \right)$.}. The expression of ${\boldsymbol\vartheta}^{ab}$ imposes a constraint on ${\boldsymbol\lambda}^{ab}$ as
\begin{align}\label{lam expression}
    \boldsymbol\lambda^{ab}  = {\left( \boldsymbol\vartheta  \right)^{ - 1}}{^{ac}}\delta {\boldsymbol\vartheta _{c}}^b.
\end{align}

Based on the new variation $\boldsymbol{\delta}$ and the expression of $\boldsymbol{\lambda}$ in \eqref{lam expression}, we get an improved definition of the corner symplectic potential,
\begin{eqnarray}\label{equivalence of sym}
   {\bf\Theta}^{{\cal S}} &&:= {\left. {\left(V\delta \boldsymbol{q}- { {{\bf{B}}^{\mu a}}{{\boldsymbol\delta }}{{{\bf{ e}}}_{\mu a}}} \right)} \right|_{\cal S}}\label{new symplec cons} \nonumber \\
   &&={\left. {\left( {\boldsymbol{p}\delta \boldsymbol{q} - {{\bf B}^{T\mu }}\delta {{\bf e}^T_{\mu a}}-{{\bf B}^{S a}}\delta {{\bf e}^S_{\mu a}}} \right)} \right|_{\cal S}}=\Theta^{c}~~\text{in \eqref{cononical sympl}}.
\end{eqnarray}
The relation \eqref{equivalence of sym} indicates that obtaining the symplectic potential \eqref{cononical sympl} in Maurer-Cartan form requires to constrain the form of $V$ in $\phi_3$ at the corners. Note that $V$ in \eqref{equivalence of sym} is self-consistent with the GaEMs, $\left({{\bf B}^{T\mu a}}, {{{\bf e}^{\mu a}},{\boldsymbol \vartheta}^{ab}({\boldsymbol p}, {\boldsymbol q})}\right)$, because the $V$ considered in \eqref{phi3 cons} is indeed a function of ${\bf e}^{\mu a},{\bf B}^{\mu a}$ and $n^a$.

{With the full symplectic potential, $\Theta^F$, derived from the variation of the action \eqref{eq:BF_expanded}, and the symplectic potential at the corners \eqref{cononical sympl}, we have the extended symplectic potential, $\Theta^{\text{ext}}$,  defined by the difference between the full potential and this corner term  as}
\begin{align}\label{eq:extended-potential}
    {\boldsymbol\Theta}^{\text{ext}} := \Theta^{F} - {\boldsymbol\Theta}^{{\cal S}},
\end{align}
with $\Theta^F = \int_\Sigma (B^a e_a + B\omega)$. The presence of $\Theta^F$ here is due to the torsion-free condition, $T_a=0$ and the equations of motion correspond to the action \eqref{eq:BF_expanded} \cite{Frodden:2019ylc}. 

To ensure the gauge invariance of the bulk fields, we introduce the \textit{partial gauge transformation} $\delta_{\alpha}$ and Lorentz transformation $\delta_{\boldsymbol\lambda}$ to separate the transformations on the corner variables into two parts
\begin{subequations}\label{trans one}
\begin{align}
    % Alpha transformations (Eq. 1a)
    & \delta_{\alpha} \boldsymbol{q} = f_{\alpha} \boldsymbol{q}, \quad \delta_{\alpha} V = -f_{\alpha} V, \quad \delta_{\alpha} \mathbf{e}^{\mu a} = \delta_{\alpha} \mathbf{B}^{\mu a} = 0, \qquad \delta_{\alpha} \boldsymbol{\vartheta} = - \alpha \boldsymbol{\vartheta}, \quad \delta_{\alpha} \boldsymbol{\vartheta}^{-1} = \boldsymbol{\vartheta}^{-1} \alpha, \label{corner transfor} \\
    % Lambda transformations (Eq. 1b)
    & \delta_{\boldsymbol\lambda} \mathbf{e}^{\mu a} = \boldsymbol\lambda^{ab} \mathbf{e}^{\mu}_{b}, \quad \delta_{\boldsymbol\lambda} \mathbf{B}^{\mu a} = \boldsymbol\lambda^{ab} \mathbf{B}^{\mu}_{b}, \qquad \delta_{\boldsymbol\lambda} \boldsymbol{q} = \delta_{\boldsymbol\lambda} V = \delta_{\boldsymbol\lambda} \boldsymbol{\vartheta} = \delta_{\boldsymbol\lambda} \boldsymbol{\vartheta}^{-1} = 0. \label{lambda transfor}
\end{align}
\end{subequations}
Here, we introduce $f_\alpha$ to parameterize the transformation of $\boldsymbol{q}$ and $V$ under $\alpha$, which means $f_{\alpha}$ can be directly interpreted as the rescaling factor for $\boldsymbol{q}$. This also indicates that the partial gauge transformation is in fact a combination of a rescaling transformation and a Lorentz transformation. The covariant variation $\boldsymbol{\delta}$ represents the physical change in the fields. The compensating term $\delta_{\boldsymbol{ \lambda}}$ is introduced specifically to cancel the pure gauge degrees of freedom that arise from the GaEMs. A similar procedure was also  discussed in \cite{Freidel:2020svx,Donnelly:2016auv,Gomes:2016mwl}.

In this sense, according to \eqref{new symplec cons}, the gauge part $\delta_{\boldsymbol{\lambda}}$ is identified with the external geometric quantities and the contribution of the $B$-field within $\Theta^c$ (characterized by  ${\boldsymbol{p}}\delta {\boldsymbol{q}}$). This offers a novel perspective to understand the role of the external geometry at the corners. We aim to verify that this formalism indeed satisfies the expected properties, particularly with respect to the integrability of the charge and the explicit classification of configuration-preserving transformations of the corners.
To this end, we decompose the total symplectic structure into a part on $\Sigma$ and a part on $\cal S$:
\begin{align}
    {\boldsymbol\Omega}^{\text{ext}}=\delta {\boldsymbol\Theta}^{\text{ext}}=\Omega_\Sigma-\Omega_{\cal S}=\int_\Sigma  { \left( {\delta {B^a}\delta {e_a} + \delta B\delta \omega } \right)}  + {\left. {\delta\left( {  {{\bf B}^{\mu a}}\boldsymbol\delta {{\bf e}_{\mu a}}} \right)} \right|_{\cal S}}.
\end{align}
By contracting the symplectic structure ${\boldsymbol\Omega}^{\text{ext}}$ with the operator $\delta_{\alpha}\lrcorner$, we can obtain the charge $Q$:
\begin{align}\label{charge definition}
    \begin{aligned}
 \slashed{\delta} Q \equiv -\delta_\alpha \lrcorner {{\boldsymbol\Omega}^{\text{ext}} }= -\delta_\alpha \lrcorner {\Omega_{\Sigma} } +\delta_\alpha \lrcorner {\Omega_{\cal S} },
    \end{aligned}
\end{align}
where $\slashed{\delta}$ indicates that the charge $Q$ is not necessarily integrable. Gauge invariance implies that after subtracting the pure gauge part from the symplectic potential in \eqref{eq:extended-potential}, the gauge charge associated with the remainder, ${\boldsymbol\Omega}^{\text{ext}}$, vanishes.

Let us now calculate the two terms in the r.h.s. of \eqref{charge definition}, $ -\delta_\alpha \lrcorner {\Omega_{\Sigma} }$ and $ -\delta_\alpha \lrcorner {\Omega_{\cal S} }$, respectively. Using the bulk fields transformations in \eqref{bulk field trans}, the first term is
\begin{align}
    -\delta_\alpha \lrcorner {\Omega_{\Sigma} }={\left. {\delta \left( {  {\alpha _{IJ}}{B^{IJ}}}\right)} \right|_{\cal S}},
\end{align}
where ${\alpha _{IJ}} = {\epsilon _{IJ}}^K{\alpha _K}$ and ${B^{IJ}} = {\epsilon _{IJ}}^K{B_K}$ satisfy ${\alpha _{IJ}}{B^{IJ}} = \alpha_{ab} B^{ab} + {\alpha _a}{\tilde B^a}$. The raising and lowering of indices are now performed using $\delta^{IJ}$ from \eqref{internal metric}. For the second term, due to the transformations of the corner variables in \eqref{corner transfor}, we get
\begin{align}
    \delta_\alpha \lrcorner \Omega_{\cal S} = \left. \left[\delta_\alpha \left( \boldsymbol\vartheta^{-1} \boldsymbol\lambda \boldsymbol\vartheta \right)_{ab} {\bf B}^{ab}-{\delta _\alpha }\left( {V\delta {\boldsymbol{q}}} \right)- \delta \left(f_\alpha V\boldsymbol{q}- \left( \boldsymbol\vartheta^{-1} \alpha \boldsymbol\vartheta \right)_{ab} {\bf B}^{ab} \right)  \right] \right|_{\cal S},
\end{align}
where the term beyond $\delta(\cdots)$ can be calculated as
\begin{align}
      {\delta _\alpha }{\left( {{\boldsymbol\vartheta ^{ - 1}}\boldsymbol\lambda \boldsymbol\vartheta } \right)_{ab}}{{\bf B}^{ab}} =   {\left[ {\alpha ,\boldsymbol\lambda } \right]_{ab}}{\bf B^{ab}}.
\end{align}
Since the transformation rule introduced in \eqref{corner transfor} ensures ${\delta _\alpha }\left( {V\delta {\boldsymbol{q}}} \right)=0$,  we get that the charge $Q$ on $\cal S$ is integrable if
\begin{align}\label{integrable condition}
    \left[ {\alpha ,\boldsymbol\lambda } \right]_{01}=0,\quad \text{or equivalently}\quad \alpha_{00}=\alpha_{11}.
\end{align}
Because this condition is automatically satisfied, we obtain a well-defined expression for the charge $Q$:
\begin{align}\label{expression of charge}
    Q=  {\alpha _{IJ}}{B^{IJ}} +f_\alpha V\boldsymbol{q}-{\left( {{\boldsymbol\vartheta ^{ - 1}}\alpha \boldsymbol\vartheta } \right)_{ab}}{{\bf B}^{ab}},
\end{align}
from which we have a gluing condition 
\begin{align}\label{glueing condition}
     {\alpha _{ab}}{B^{ab}} + {\alpha _0}{\tilde B_0} + {\alpha _1}{\tilde B_1} \mathop  = \limits^{\cal S}  {\left( {{\boldsymbol\vartheta ^{ - 1}}\alpha \boldsymbol\vartheta } \right)_{ab}}{{\bf B}^{ab}}-f_\alpha V\boldsymbol{q},
\end{align}
fulfilling $Q=0$.
Without loss of generality, assuming $\alpha_0=\alpha_1=0$, the gluing condition can be rewritten in a form that is dependent on $(\alpha^{-1})^{ab}$ and $f_\alpha$ as
\begin{align}
  B^{ab} \overset{\mathcal{S}}{=}  \left( \boldsymbol{\vartheta} \mathbf{B} \boldsymbol{\vartheta}^{-1} \right)^{ab}-\frac{1}{2}(\alpha^{-1})^{ab}f_\alpha V\boldsymbol{q}.
\end{align}
In \cite{Freidel:2020svx}, the Maurer-Cartan form $\boldsymbol{\lambda} = \boldsymbol{\vartheta}^{-1}\delta\boldsymbol{\vartheta}$ naturally facilitates the use of left and right transformations in \eqref{corner transfor}. This mechanism is crucial, as it ensures the constraint algebra, specifically the Poisson bracket of their corner charge ${\cal C}_{\rm{BF}}$, is closed, i.e., $\{{\cal C}_{\rm{BF}}, {\cal C}_{\rm{BF}}\} \sim {\cal C}_{\rm{BF}}$. This principle can be generalized to the charge $Q$ in our work.

So far, we have obtained the conditions for gauge invariance, and now we wish to restore the degrees of freedom of the system at the corners. To this end, we introduce another partial gauge transformation
\begin{align}\label{trans two}
    {{\bar \delta }_{{\bar\alpha}}}{{\bf e}^{\mu a}} = -{\bar\alpha ^{ab}}{{\bf e}^\mu }_b,\quad{{\bar \delta }_{{\bar\alpha}}}{{\bf B}^{\mu a}} = {\bar\alpha ^{ab}}{{\bf B}^\mu }_b,\quad{{\bar \delta }_{\bar\alpha}}{\boldsymbol q}={{\bar \delta }_{\bar\alpha}}{ V}={{\bar \delta }_{\bar\alpha}}{\boldsymbol\vartheta ^{ab}} = 0.
\end{align}
We can classify the transformation $\bar \delta_{\bar\alpha}$ into nine generators of the $\mathfrak{sl}(2,\mathbb{R})$ algebra mentioned in Section \ref{subsec:Corner algebra}. This choice is natural because ${\bf e}^{\mu a}, {\bf B}^{\mu a}$ and ${ \boldsymbol{q}}$ are decoupled. Consequently, the transformations generated by these nine generators all act trivially on $\boldsymbol{\vartheta}$, which is related to $\boldsymbol{q}$.  In this case, the Poisson bracket can be written in a form involving the matrix $\bar \alpha^{ab}$, for example, ${\left\{ {J_0^{(T)},{{\bf e}^{\mu a}}} \right\}_P} = \frac{1}{2}\delta _T^a{n^b}{{\bf e}^{\mu T} }$ for $\bar\alpha^{ab}=\frac{1}{2}\delta _T^a{n^b}$.

By contracting $\Omega^{\text{ext}}$ with $\bar \delta_{\bar\alpha} \lrcorner$, we obtain the integrable charge as
\begin{align}
 -\delta_{\bar\alpha} \lrcorner  \Omega^{\text{ext}}  = \delta {\cal Q} =  - \delta \left( {{\bar\alpha _{ab}}{{\bf B}^{\mu a}}{{\bf e}_\mu }^b} \right),
\end{align}
which corresponds to the charges related to the degrees of freedom at the corners. {It is worthwhile to emphasize that besides the symmetries generated by the generators of the $\mathfrak{sl}(2,\mathbb{R})$ algebra  for the field $\boldsymbol{\vartheta}$ we previously discussed, there are also two other classes of symmetries.} The first is a conjugate transformation generated by the generator $\boldsymbol g$ of the ${\mathfrak{so}}(2)$ algebra, described by the relation $\boldsymbol{\vartheta}\to \boldsymbol g^{-1}\boldsymbol{\vartheta}\boldsymbol g$, which is a trivial transformation in this type of algebra. The second is a rescaling of $y^a$ and $n^a$: $y^a\to\frac{1}{\lambda_r}y^a$ and $n^a\to\lambda_r n^a$, which does not change $\boldsymbol{p}$ and $\boldsymbol{q}$ based on \eqref{eq:canonical_vars}.

The charges we have derived from different physical symmetries constitute non-trivial degrees of freedom at the corners, and gauge symmetry is also successfully maintained. Moreover, we argue that the function $V$, which determines the wiggling boundary information, determines not only the observables according to \eqref{V form} but also determines whether the extrinsic vectors can be packaged into the Maurer-Cartan form. It is worth noting that the transformations generated by $\bar \delta_{\alpha}$ in \eqref{trans two} differ from the approach in \cite{Donnelly:2016auv,Speranza:2017gxd,Setare:2018mii}. Instead of the surface transformations acting on corner variables, we consider the transformations generated by the observables we introduced in Section \ref{subsec:Corner algebra}.

\section{Conclusions and discussions}
\label{sec:conclusions}
In this paper, we investigated the GrEMs and the GaEMs for solutions of JT gravity with defects, as well as the algebras and charges of these modes at corners.

 Firstly, we focused on the GrEMs corresponding to solutions for conical defect and wormhole within the second-order formulation. For the conical defect, {we employ the generalized F-G gauge, and introduced  compatible internal symmetries for the GrEM, $\theta$ , by constraining the topology to maintain a constant conical defect angle.} We found that the boundary dynamics are subsequently governed by the Schwarzian action in terms of the GrEM. However,  the dynamics are  modified by the inclusion of terms related to the black hole temperature and the radial displacement, and when their values  at the horizon in the base spacetime are constant,   the boundary action retains both $\rm{SL}(2, \mathbb{R})$ symmetry and temperature rescaling invariance.  These findings provide a description of how physical quantities in the bulk serve to break the boundary symmetry. Furthermore, the ultraviolet and infrared behaviors of the boundary action are found to be determined respectively by higher-order perturbations involving these temperature and radial displacement-related terms, and by local effects.

For the wormhole defect solution, we likewise derive constraints on the GrEM, $\tilde{\theta}$, via the transformation from base to target spacetimes, which was found to characterize the variations in the circumference of the wormhole throat. The corresponding boundary action in this case manifests as a deformed Schwarzian theory. Significantly, in the limit where the dilaton field approaches a constant and the variation in the wormhole throat circumference is small, this deformed theory was found to admit a physical interpretation as describing a particle with time-dependent mass subject to a constant external force.  Additionally,  we found that the presence of the wormhole precludes the simultaneous preservation of both time-translation and $\phi$-translation symmetries.

Then, we investigated the GaEMs at corners.  We found that the difference between the first- and second-order formulations of JT gravity on a codimension-one surface can be cancelled by a residual dilaton tensor, which localizes their discrepancy to the corners.   Additionally,  we found that the corner configuration can also be  preserved under certain constraints  based on  the relationship between tangential and normal vectors, which can be self-consistently constructed by forming canonical pairs and reduces the Poisson brackets of these canonical pairs to Dirac brackets. In addition to the configuration-preserving constraints, we introduce another constraint corresponding to the wiggling boundary information. This not only renders the system at the corners a system with gauge redundancy but also links the GrEMs and the GaEMs.

In particular, the canonical pairs we identify directly construct the generators of an $\mathfrak{sl}(2,\mathbb{R})$ algebra. Crucially, we introduced a novel partitioning of the fields into transverse and longitudinal components, which provides a systematic and detailed classification of this generator algebra. And we demonstrated that under configuration- and boundary-information-preserving constraints, the corners are  first-class constrained systems, in which  we managed to  define the observables, saying the fundamental transformations corresponding to physical symmetries, entirely dictated by wiggling boundary information. A subsequent key finding is that, in a unitary representation, the Casimir operators, in their form as the discrete square of the ``parallelogram area'', reveal an invariance among the fields on the corner in the sense of a ``geometric quantity''.  Further classifying these transformations yields the corner charges, thereby establishing a one-to-one correspondence between boundary information and the physical degrees of freedom. Moreover, we systematically derived the GaEMs  by isolating the pure gauge part from the symplectic potential, and achieve the gauge invariance by compactly packaging the extrinsic vectors into the Maurer-Cartan form. And finally, we successfully constructed the corner charges under the gauge invariance condition, as well as the integrability of the charge in JT gravity.

Along with our study, there are still some directions which deserve further exploration. First, the boundary action in \eqref{boundary action main}, takes the GrEM, $\theta$, as the integration variable, deserves further investigation, because  its generalization to non-constant temperature and horizon location could possess a richer structure. One way is to compute its associated correlation functions by considering  an expansion near the $\theta=t$ point where the wiggling effect tends to vanish. Note that the calculation of correlation functions cannot be performed as in previous works \cite{Botta-Cantcheff:2018brv, Botta-Cantcheff:2019apr,Arias:2021ilh}, as the physical quantities are time-dependent in this case. Additionally, the effect of dimensional reduction on the gluing condition of GaEMs can be considered, following the approach in \cite{Joung:2025kin}. Another interesting  direction is using the canonical pairs at the corners to construct generators for other symmetry groups, such as the $\rm{SL}(3, \mathbb{R})$ group in \cite{Ozer:2025bpb}, the $\rm{SO}(2, 2)$ group in \cite{Chirco:2024ubu} and the universal corner symmetry in \cite{Ciambelli:2024qgi}.  The last but not the least, it could be interesting to further explore the physical meaning of the parallelogram-type Casimir operators on the asymptotic boundary, and generalize these operators to other types of boundaries. For instance, a comparison could be made with light-like boundaries, and also with the method in \cite{Wieland:2017cmf} which obtains invariants using null boundaries and spinor variables.

\section*{Acknowledgment}
This work is partly supported by the Natural Science Foundation of China under Grants No. 12375054 and the Postgraduate Research $\&$ Practice Innovation Program of Jiangsu Province under Grants No. KYCX-3502.

	\appendix
	\section{Notation and conventions}
	\label{appe:appendix one}
	In this appendix, we shall clarify some notations and conventions of our calculations. We choose $\mu,\nu,\alpha\in \left\{ {0,1} \right\}$,   $a,b,c,\cdots \in \left\{ {0,1} \right\}$ and $I,J,K,\cdots \in \left\{ {0,1,X} \right\}$ as coordinate indices and abstract indices, respectively. We define $\eta^{ab}$,  $ \tilde\eta^{ab}$,  $ \delta^{IJ}$ as the internal metrics in the form

    \begin{align}\label{internal metric}
	{\eta ^{ab}} = \left( {\begin{array}{*{20}{c}}
			{ -1}&\\
			&1
	\end{array}} \right),\quad\quad {{\tilde \eta }_{ab}} = \left( {\begin{array}{*{20}{c}}
{  1}&0\\
0&1
\end{array}} \right),\quad\quad {\delta ^{IJ}} = \left( {\begin{array}{*{20}{c}}
1&{}&{}\\
{}&1&{}\\
{}&{}&1
\end{array}} \right).
	 \end{align}
	For a $ d$-dimensional spacetime, the expression adopted for an arbitrary $p$-form is
	\begin{align}
		P=\frac{1}{p!} P_{\mu_1 \ldots \mu_p} \mathrm{~d} x^{\mu_1} \wedge \cdots \wedge \mathrm{~d} x^{\mu_p}.
	\end{align}
The induced metric of the considered timelike boundary satisfies:
\begin{align}
    {h_{\mu \nu }} = {g_{\mu \nu }} + {n_\mu }{n_\nu }
\end{align}
and its adapted derivative operator is $D_\mu$.

We define the Levi-Civita symbol $ \varepsilon_{\mu\nu}$ satisfying $\varepsilon_{01}=1$. Using $\varepsilon_{\mu\nu}$, a two-form is defined as
	\begin{align}
		\epsilon:=\frac{1}{2}\epsilon_{\mu\nu}dx^{\mu}\wedge dx^{\nu},
	\end{align}
where the Levi-Civita tensor $\epsilon_{\mu\nu}:=\sqrt { - g} \varepsilon_{\mu\nu}$ satisfies
	\begin{align}
		{\epsilon ^{\mu \nu }} = {g^{\mu \alpha }}{g^{\nu \beta }}{\epsilon _{\alpha \beta }} = {g^{\mu \alpha }}{g^{\nu \beta }}\sqrt {|g|} {{ \varepsilon }_{\alpha \beta }} = {g^{ - 1}}\sqrt {|g|} {{ \varepsilon }_{\mu \nu }}.
	\end{align}
  
Define the Hodge dual $\star$ to satisfy $\star^2 P=-(-1)^{p(d-p)+1} P$, and
	\begin{align}\label{Hodge dual}
		\begin{aligned}
			&(\star P)^{\mu_{1}...\mu_{n-p}}  :=\frac{1}{p!}P_{\alpha_{1}..\alpha_{p}}\epsilon^{\alpha_{1}..\alpha_{p}\mu_{1}..\mu_{n-p}}, \\
			&P_{\alpha_{1}..\alpha_{p}}  :=-\frac{1}{(n-p)!}\epsilon_{\alpha_{1}..\alpha_{p}\mu_{1}..\mu_{n-p}}(\star P)^{\mu_{1}..\mu_{n-p}}.
		\end{aligned}
	\end{align}
	
By contracting $\epsilon_{\mu\nu}$ with ${e_\mu}^a$, one can obtain
		\begin{align}
			\varepsilon_{ab}=e^{\mu}{}_{a}e^{\nu}{}_{b}\epsilon_{\mu\nu}.
		\end{align}
	Combining with the definition of the Hodge dual \eqref{Hodge dual}, the one-form $\alpha_\Sigma$ and the zero-form $\alpha_{\cal S}$ can be expressed as
	\begin{align}
{\alpha _\Sigma } = {({\star\alpha _\Sigma })^\mu }{\epsilon _{\mu \nu }}{\rm{d}}{x^\nu }\label{partial integral one}
	\end{align}
	where
		\begin{align}\label{dual relation}
			{(\star{\alpha _\Sigma })^\mu } = {\alpha _\Sigma }_\nu {\epsilon ^{\mu\nu }}.
		\end{align}
With $n^\mu$ and $y^\mu$ serving as the respective normal vectors for the surfaces $\Sigma$ and $\cal S$, the integrals of $\text{d}\alpha_\Sigma$ and $\text{d}\alpha_{\cal S}$ over $\cal M$ and $\Sigma$, respectively, yield
\begin{align}
			\int_{\cal M }{{\rm{d}}{\alpha _\Sigma }}  = \int_{\cal M} {\sqrt { - g} {\nabla _\mu }{{( \star {\alpha _\Sigma })}^\mu }}  = \int_\Sigma  {{\alpha _\Sigma }}  = \int_\Sigma  {\sqrt { - h} {{( \star {\alpha _\Sigma })}^\mu }{n_\mu }}.
		\end{align}
It can be verified that the two types of frames, ${e_\mu}^a$, as well as the tensors $\epsilon_{\mu\nu}$ and $\epsilon_{ab}$, satisfy the relations
		\begin{align}\label{two key relations}
			\varepsilon_{ab} \epsilon^{\mu\nu}=2  e^{[\mu}{ }_a e^{ \nu}{}_b{ }^{]}.
		\end{align}

\section{Examining the relationship between the first-order and second-order symplectic potentials}
	\label{appendix: Examining the relationship between the first-order and second-order symplectic potential}

This appendix facilitates a more direct understanding of the role the corner term plays in the equivalence between the first- and second-order formulations, namely, the explicit verification of relation \eqref{eq:new relation}.

	The Hodge dual of the symplectic potential in \eqref{eq:first order symplectic} can be rewritten using the definition of the spin-connection:
	\begin{align}
		\begin{aligned}\label{dual first}
		&{( \star {\theta _f})^\mu } =\frac{1}{2} {B}{_{ab}}\delta {\omega _\nu }^{ab}{\epsilon^{\mu \nu }} =   {B} e^{[\mu}{ }_a e^{ \nu}{}_b{ }^{]}\delta {\omega _\nu }^{ab} =2 {B}{e^{[\mu }}_b{e^{\nu ]}}_a(\delta {e_\alpha }^a{\nabla _\nu }{e^{\alpha b}} + {e_\alpha }^a{\nabla _\nu }\delta {e^{\alpha b}} + {e_\alpha }^a\delta \Gamma _{\nu \beta }^\alpha {e^{\beta b}})\\
		&={B}( \star{\theta _{EH}})^\mu + 2{B}{\nabla _\nu }({e^{[\nu }}_a\delta {e^{\mu] a}}) ={B}(\star{\theta _{EH}})^\mu + 2{\nabla _\nu }({B}{e^{[\nu }}_a\delta {e^{\mu] a}}) -2 {e^{[\nu }}_a\delta {e^{\mu] a}}{\partial _\nu }{B},
		\end{aligned}
	\end{align}
	where \eqref{two key relations} and \eqref{dual relation} have been used, and
	\begin{align}
		( \star{\theta _{EH}})^\mu  =  {{\nabla _\nu }{{(\delta g)}^{\mu \nu }} - {\nabla ^\mu }(\delta g)_\nu ^\nu }.
	\end{align}
The result \eqref{dual first} shows that the total derivative term cancels the part containing derivatives of $\delta g_{\mu\nu}$ from the second-order formalism, such that the first-order symplectic potential depends only on $\delta \omega^{ab}$. According to \cite{Goel:2020yxl, Arias:2021ilh}, the integral over $\Sigma$ of the first term on the r.h.s of the final equality in \eqref{dual first} can be written as
	\begin{align}\label{dual relation}
	\int_\Sigma  {\sqrt {\left|h \right|} {B}{n_\mu }(\star\theta _{EH} )^\mu}  = \int_{\Sigma} {\sqrt {\left| h \right|} } {B}{D_\mu }(\delta {n^\mu }+g^{\mu\nu}\delta n_\nu) - \delta \left( 2{\int_\Sigma  {\sqrt {\left| h \right|} {B}K} } \right) +2 \int_\Sigma  {\sqrt {\left| h \right|} \delta {B}K}.
	\end{align}
in which the first term on the r.h.s can be rewritten as
\begin{align}
    \int_{\Sigma} {\sqrt {\left| h \right|} } {B}{D_\mu }(\delta {n^\mu }+g^{\mu\nu}\delta n_\nu) =\int_{\Sigma} {\sqrt {\left| h \right|} } {D_\mu }({B}(\delta {n^\mu }+g^{\mu\nu}\delta n_\nu) )-\int_{\Sigma} {\sqrt {\left| h \right|} } (\delta {n^\mu }+g^{\mu\nu}\delta n_\nu) D_\mu{B}.
\end{align}
Combining the above formula and \eqref{second order sym} yields the relation \eqref{eq:second and first relation}.

Furthermore, the final term in \eqref{dual first} can be rewritten as
 \begin{align}\label{extra term}
 \begin{aligned}
&-{n_\mu }{e^{[\nu }}_a\delta {e^{\mu] a}}{\partial _\nu }{B}\\
& = \frac{1}{2}{n_\mu }{e^{\mu a}}{\partial _\nu }{B}\delta {e^\nu }_a - \frac{1}{2}{e^\nu }_a{n_\mu }{\partial _\nu }{B}\delta {e^{\mu a}}\\
& = \frac{1}{2}{n^a}{\partial _\nu }{B}{g^{\alpha \nu }}\delta {e_{\alpha a}} -\frac{1}{2} {n^\alpha }{\partial _a}{B}\delta {e_\alpha }^a.
 \end{aligned}
 \end{align}
To evaluate the relevant terms, we use the following two relations
\begin{subequations}
\begin{align}
&{g^{\mu \nu }}\delta {g_{\nu \alpha }}{\partial ^\alpha }{B} = {e^{\mu I}}\delta {e_{\alpha a}}{\partial ^\alpha }{B} + {g^{\mu \nu }}\delta {e_{\nu a}}{e_{\alpha a}}{\partial ^\alpha }{B},\\
&{\partial ^\mu }{B}{g^{\alpha \nu }}\delta {g_{\alpha \nu }} = 2{\partial ^\mu }{B}{e^{\alpha a}}\delta {e_{\alpha a}}.
 \end{align}
\end{subequations}
These two results lead to the following relation
\begin{align}\label{eq:B7}
\begin{aligned}
{n_\mu }{e^{[\nu }}_a\delta {e^{\mu ]a}}{\partial _\nu }{B} &= \frac{1}{2}{n_\mu }\left( {{g^{\mu \nu }}\delta {g_{\nu \alpha }}{\partial ^\alpha }{B} - {g^{\mu \nu }}\delta {e_{\nu a}}{\partial ^a }{B}} \right) - \frac{1}{2}{n^\alpha }{\partial _a}{B}\delta {e_\alpha }^a\\
 &=n^\alpha {\partial_\alpha }{B}{e^{\mu a}}\delta {e_{\mu a}} - {n^\alpha }{\partial _a}{B}\delta {e_\alpha }^a - \frac{1}{2}\left( {{n^\alpha }{\partial _\alpha }{B}{g^{\mu \nu }}\delta {g_{\mu \nu }} - {n^\nu }\delta {g_{\nu \alpha }}{\partial ^\alpha }{B}} \right)\\
& = {{\cal T}^{\alpha a}}\delta {e_{\alpha a}} - \frac{1}{2}\left( {{n^\alpha }{\partial _\alpha }{B}{g^{\mu \nu }}\delta {g_{\mu \nu }} - {n^\nu }\delta {g_{\nu \alpha }}{\partial ^\alpha }{B}} \right),
\end{aligned}
    \end{align}
Subsequently, we combine  \eqref{dual first},
\eqref{dual relation}, \label{eq:B7} and
\begin{align}
    {n^\alpha }{\partial _\alpha }{B}{g^{\mu \nu }}\delta {g_{\mu \nu }} - {n^\nu }\delta {g_{\nu \alpha }}{\partial ^\alpha }{B} = {n^\alpha }{\partial _\alpha }B{h^{\mu \nu }}\delta {h_{\mu \nu }} + \left( {\delta {n^\alpha } + {g^{\alpha \beta }}\delta {n_\beta }} \right){D_\alpha }B
\end{align}
shall give the expression \eqref{second order sym}.

%Additionally, the specific contribution to the symplectic potential from ${{\cal T}^{\alpha a}}$, which is the part on $\Sigma$ containing the zweibein, can be seen by examining different specific solutions.
%At the asymptotic boundary of $\rm{AdS}_2$ in \eqref{target ads}, the following relations hold
%\begin{subequations}\label{approxiamte relations}
   %%%&{n^\alpha }{e^{\beta a}}{\partial _\beta }{B}\delta {e_{\alpha a}} = {n^\tau }{e^{ra}}{\partial _r}{B}\delta {e_{\tau a}} + {n^\tau }{e^{\tau a}}{\partial _\tau }{B}\delta {e_{\tau a}},
   %%% \end{align}
%\end{subequations}
%which imply ${{\cal T}^{\alpha a}}\delta {e_{\alpha a}}={\cal O}\left( {e^{r}} \right)$. This indicates that ${{\cal T}^{\alpha a}}\delta {e_{\alpha a}}$ has a non-zero contribution on the asymptotic boundary due to the non-vanishing variations of functions ${\cal L}^0$, ${\cal L}^-$ and ${\cal L}^+$. Since $\delta {e_{\alpha a}}=0$ for every component of the corresponding the zweibein at the boundary $z\to 0$ of the wormhole solution \eqref{wormhole solution}, the deviation between the symplectic potentials in first- and second-order formulation, ${{\cal T}^{\alpha a}}\delta {e_{\alpha a}}$, vanishes. However, this result depends on the coordinate transformations, because in the framework of the wiggling boundary, the variations of $\tilde w$ and $\tilde \theta$ are generally not zero.

%\bibliography{ref}
%\bibliographystyle{utphys}
\providecommand{\href}[2]{#2}\begingroup\raggedright\endgroup

\end{document}